\documentclass[pre,twocolumn,preprintnumbers,showpacs,floatfix]{revtex4}
\usepackage{graphicx} 
\usepackage[toc,page]{appendix}
\usepackage{subfig}
\usepackage{amsmath}
\usepackage{amssymb}
\usepackage{multirow}
\usepackage{url}

\usepackage{color,soul}
\begin{document}

\title{Scaling Properties of Force Networks for Compressed Particulate Systems}
\author{L. Kovalcinova, A. Goullet, and L. Kondic}
\affiliation{
Department of Mathematical Sciences,
New Jersey Institute of Technology,
University Heights,
Newark, NJ 07102\\
}

\date{\today}

\begin{abstract}
We consider, computationally and experimentally, the scaling properties of force networks in the systems of circular particles 
exposed to compression in two spatial dimensions.   The simulations
consider polydisperse and monodisperse particles, both frictional and frictionless, and in experiments we use monodisperse and 
bidisperse particles.   While for some of the considered systems we observe 
consistent scaling exponents describing  the behavior of the force networks, we find that this behavior is {\it not universal}.  In particular, 
monodisperse frictionless systems that partially crystallize 
under compression, show scaling properties that are significantly different compared to the other considered systems. The findings of 
non-universality are confirmed by explicitly computing fractal 
dimension for the considered systems.  The results of physical experiments are consistent with the results obtained in simulations of 
frictional particles.
\end{abstract}


\maketitle

\section{Introduction}

\begin{figure}[t]
\centering
 \subfloat[]{\hspace{-0.0cm}\includegraphics[width=1.7in]{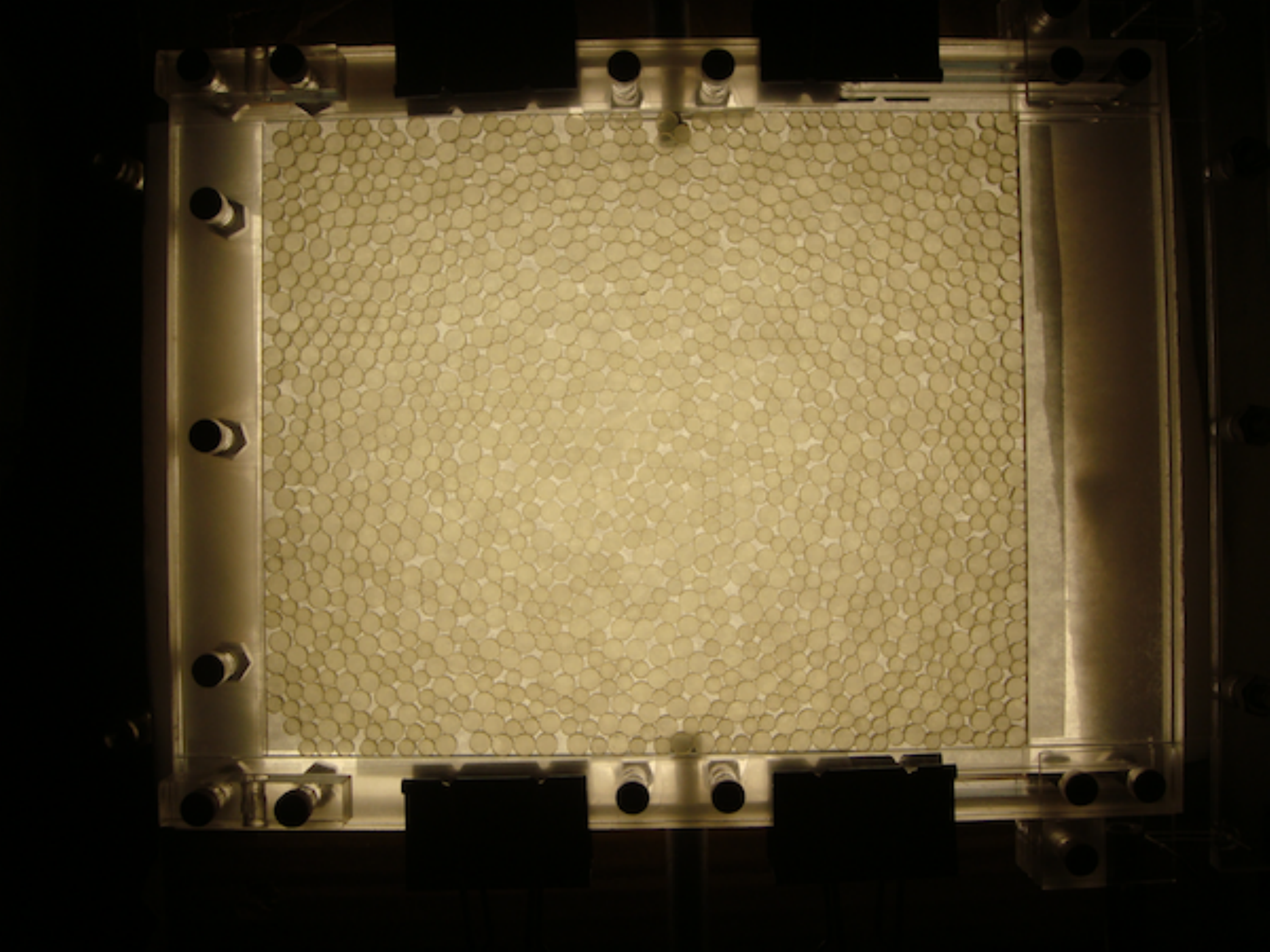}}
 \subfloat[]{\includegraphics[width=1.7in]{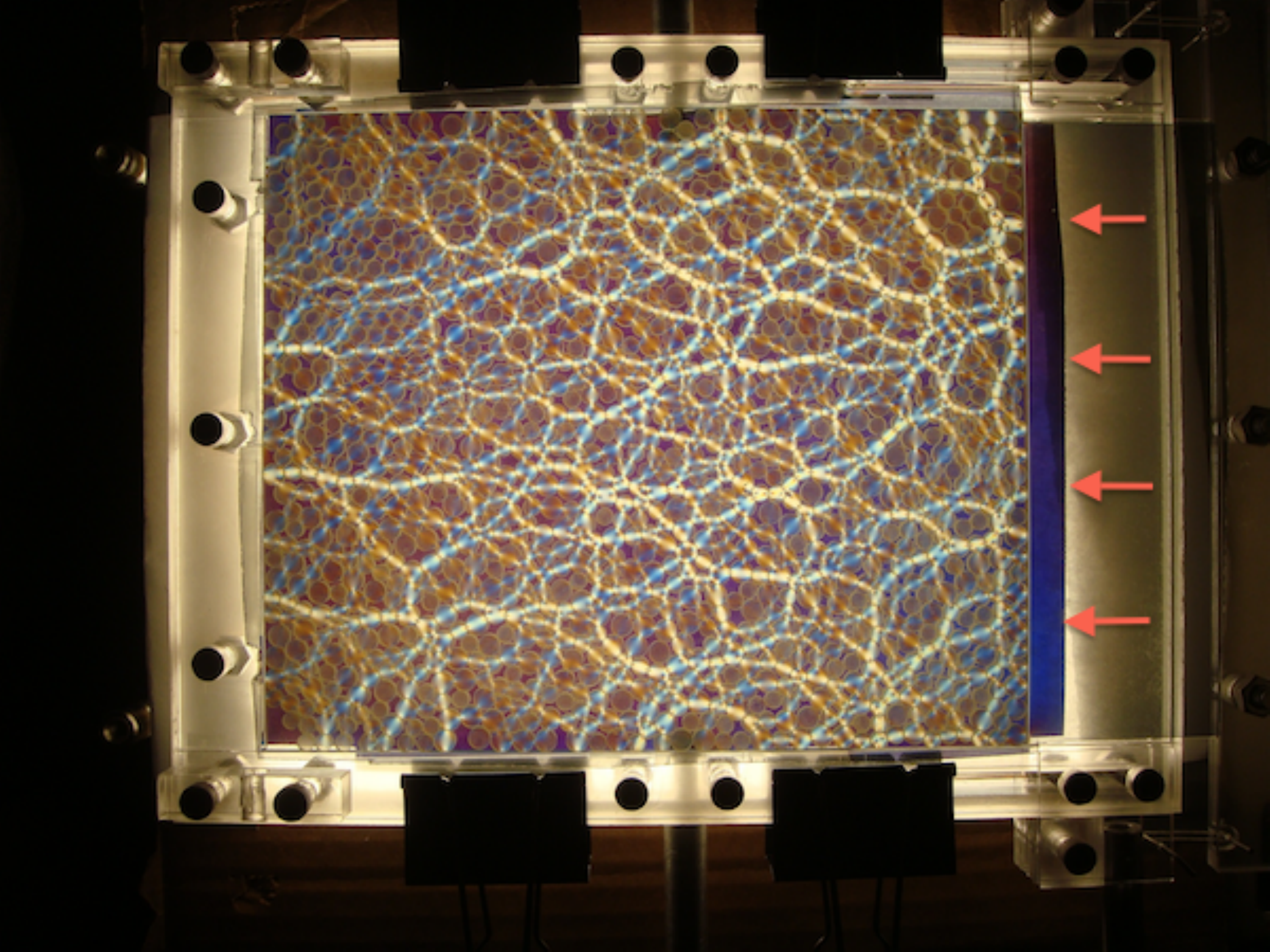}}
\caption{Experimental images of a two-dimensional system of photoelastic particles, obtained (a) without and (b) with cross-polarizer. 
The arrows indicates the side of the container where pressure is applied.}\label{exp}
\end{figure}

Particulate materials are relevant in a variety of systems of practical relevance.   It is well known that macroscopic properties of these systems are related to 
the force networks - the mesoscale structures that characterize the internal stress distribution.  The force networks are built on top of the contact networks formed
by the particles.  These contact networks have been studied using a number of approaches, see~\cite{albert_barabasi_02, alexander_physrep05} 
for reviews.   However, the properties of force networks, that form 
a subset of the contact networks based on the interaction force, are not slaved to the contact one: a single contact network can support infinite set of possible force
networks, due to indeterminacy of the interaction forces.     These force networks have been analyzed using a variety of approaches, including distributions
of the force strengths between the particles~\cite{mueth98,radjai99},  the tools of statistical physics~\cite{radjai98b,majmudar05a,silbert_pre06,Wambaugh_physD_10}, 
local properties of the force networks~\cite{peters05,tordesillas_pre10,tordesillas_bob_pre12}, networks-based type of 
analysis~\cite{daniels_pre12, herrera_pre11,walker_pre12},   as well as the topology-based measures~\cite{arevalo_pre10,arevalo_pre13}.  
Of relevance to the present work are the recent results obtained using algebraic topology~\cite{epl12, pre13,pre14}
that have shown that in particular frictional properties of the particles play an important role in determining connectivity 
properties of the considered force networks.  For illustration, Fig.~\ref{exp} shows  an example of the experimental system (discussed in more 
details later in the paper), where the particles are visualized without (a) and with (b) cross-polarizers; in the part (b) force networks are clearly visible.

The recent work~\cite{ostojic} suggests that properties of these force networks are universal.  In other words, the finding is that, when properly scaled, the
distributions of force clusters (defined as groups of particles in contact experiencing the force larger than a specified threshold) collapse to a single curve.   
In~\cite{ostojic} it has been argued that this universality finding is independent of the particle properties like polydispersity and friction, or anisotropy of the 
force networks induced by shear~\cite{ostojic_pre07}.   The influence of anisotropy on the exponents describing scaling (and universality) properties of 
force networks was further considered using $q$-model~\cite{Pastor-Satorras_12}, where it was found that anisotropy may have a strong influence on
the scaling exponents.  

In the present work, we further explore the generality of the proposed universality, using discrete element simulations in the setup where 
anisotropy is not relevant.   This exploration is motivated in part by the following `thought' experiment.   
Consider a model problem of perfectly ordered monodisperse particles that under compression form a crystal-like structure.
In such a structure, each particle experiences the same total normal force, $F=1$ (normalized by the average force). 
If we choose any force threshold $\bar F\leq 1$, we obtain only one (percolating) cluster that includes all the particles. For any $\bar F > 1$, there are no particles. 
As an outcome, the mean cluster size, $S(\bar F)$, is a Heaviside function regardless of the system size.  Considering the scaling properties of the $S$-curve for 
different system sizes is therefore trivial and the scaling exponents (discussed in detail in the rest of the paper) are not well defined.   Furthermore, the fractal dimension, $D_f$, related to the scaling exponents~\cite{stauffer}, leads trivially to $D_f=D$, where $D$ is the number of physical dimensions.  Clearly, the scaling properties of 
such an ideal system are different compared to the ones expected for a general disordered system.  Therefore,  we found at least one system where `universality' does 
not hold.  

One could argue that such perfect system as discussed above is not relevant to physical setups, so let us consider a small perturbation - for example a system of particles of
the same or similar sizes, possibly frictionless, that are known to partially crystallize under compression~\cite{epl12,pre13}.   For such systems,  $S(\bar F)$ is not a Heaviside step function,
but it may be close to it.   There is an open question therefore whether these systems (that partially crystallize) still lead to `universal' force networks.   
One significant result of this work is that this is not the case.

The remaining part of this paper is organized as follows.   In Sec.~\ref{sec:sim} we describe the simulations that were carried out.   We then follow by 
the main part of the paper, Sec.~\ref{sec:results}, where we describe computations of scaling parameters in Sec.~\ref{sec:comp};  discuss the results for the scaling exponent, $\phi$, and
the fractal dimension, $D_f$, in Sec.~\ref{sec:phi}; outline the influence of  the structural properties of considered systems in Sec.~\ref{sec:structure}; discuss the influence 
of compression rate and different jamming packing fractions in Sec.~\ref{sec:discussion}, and then present the 
results for the other scaling parameters, $\nu$ and $f_c$ in Sec.~\ref{sec:other}.  We conclude this Section by presenting the results of physical experiments that were motivated by the 
computational results in Sec.~\ref{sec:exp}.  Section~\ref{sec:conclusions} is devoted to the conclusions.

\section{Simulations}\label{sec:sim}

We perform discrete element simulations using a set of circular particles confined in a square domain, using a slow-compression 
protocol~\cite{epl12,pre13,kovalcinova_15}, augmented by relaxation as described below. Initially, the system particles are placed 
on a square lattice and are given random velocities; we have verified that the results are independent of the distribution and magnitude of 
these initial velocities.   Gravitational effects are not included; we also ignore the interaction of the particles with the 
substrate in the simulations (substrate is present in the experiments as illustrated in Fig.~\ref{exp}; these effects will be discussed elsewhere).
The discussion related to possible development of spatial order as the system is compressed can be found
in~\cite{pre13}, and will be discussed in more detail later.   In~\cite{kovalcinova_15} it was shown that the 
considered systems are spatially isotropic, that is, $4$-fold symmetry imposed by the domain shape does not influence the results.

In our simulations, the diameters of the particles are chosen from a flat distribution of the width 
$r_p$. System particles are soft inelastic disks and interact via normal and tangential forces, including static friction, 
$\mu$ (as in~\cite{epl12,pre13,kovalcinova_15}). The particle-particle (and particle-wall) interactions include 
normal and tangential components. The normal force between particles $i$ and $j$ is 
\begin{eqnarray}
 {\bf F}_{i,j,n}& =&k_n x {\bf n} - \gamma_n \bar m {\bf v}_{i,j}^n\\
 r_{i,j} = |{\bf r}_{i,j}|, \ \ \ {\bf r}_{i,j} &=& {\bf r}_i - {\bf r}_j, \ \ \  {\bf n} = {\bf r}_{i,j}/r_{i,j}\nonumber
\end{eqnarray}
where ${\bf v}_{i,j}^n$ is the
relative normal velocity.  The amount of compression is $x =
d_{i,j}-r_{i,j}$, where $d_{i,j} = {(d_i + d_j)/2}$, $d_{i}$ and
$d_{j}$ are the diameters of the particles $i$ and $j$. All quantities
are expressed using the average particle diameter, {\bf $d_{ave}$}, as
the length scale, the binary particle collision time  $\tau_c = 2\pi
\sqrt{d_{ave}/(2 g k_n)}$ as the time scale, and the average particle mass,
$m$, as the mass scale.  $\bar m$ is the reduced mass, $k_n$ (in units
of ${ m g/d_{ave}}$) is set to a value
corresponding to photoelastic disks~\cite{geng_physicad03}, and
$\gamma_n$ is the damping coefficient~\cite{kondic_99}.  The
parameters entering the linear force model can be connected to
physical properties (Young modulus, Poisson ratio) as described
 e.g. in \cite{kondic_99}.

We implement the commonly used Cundall-Strack model for static
friction~\cite{cundall79}, where a tangential spring is introduced
between particles for each new contact that forms at time $T=T_0$.
Due to the relative motion of the particles, the spring length,
${\boldsymbol\xi}$, evolves as $\boldsymbol\xi=\int_{T_0}^T {\bf
 v}_{i,j}^t~(T')~dT'$, where ${\bf v}_{i,j}^{t}= {\bf v}_{i,j} - {\bf
 v}_{i,j}^n$.  For long lasting contacts, $\boldsymbol\xi$ may not
remain parallel to the current tangential direction defined by $\bf
{t}={\bf v}_{i,j}^t/|{\bf v}_{i,j}^t|$ (see,
e.g,.~\cite{brendel98}); we therefore define the corrected
$\boldsymbol\xi{^\prime} = \boldsymbol\xi - \bf{n}(\bf{n} \cdot
\boldsymbol\xi)$ and introduce the test force 
\begin{equation}
{\bf F}_{t*} =-k_t\boldsymbol\xi^\prime - \gamma_t \bar m {\bf v}_{i,j}^t
\end{equation}
where $\gamma_t$ is the coefficient of viscous damping in the tangential
direction (with $\gamma_t = {\gamma_n}$).  To ensure that the
magnitude of the tangential force remains below the Coulomb threshold,
we constrain the tangential force to be 
\begin{equation}
 {\bf F}_t = min(\mu |{\bf F}_n|,|{\bf F}_{t*}|){{\bf F}_{t*}/|{\bf F}_{t*}|}
\end{equation}
and redefine
${\boldsymbol\xi}$ if appropriate.

The simulations are carried out by slowly compressing the domain,  starting at the packing fraction $0.63$ and ending at 
$0.90$, by moving the  walls built of $L$ monodisperse particles with diameters of size $d_{ave}$ placed initially at equal distances, $d_{ave}$, from each other. 
The wall particles move at a uniform (small) inward velocity, $v_c$, equal to $v_{0}= 2.5\cdot 10^{-5}$ (in units of $d_{ave}/\tau_c$).
Due to compression and uniform inward velocity, the wall particles (that do not 
interact with each other)  overlap by a small amount. When the effect of compression rate is explored,  the compression is
stopped to allow the system to relax.  In order to obtain statistically relevant results, we simulate a large number of initial configurations (typically $120$),  
and average the results. 

 We integrate Newton's equations of motion for both the translation and rotational
degrees of freedom using a $4$th order predictor-corrector method with
time step $\Delta T =0.02$.    Our reference system is characterized by $k_n = 4\cdot 10^3$, $e_n = 0.5$, $\mu = 0.5$, and $k_t = 0.8k_n$~\cite{goldhirsch_nature05}; 
the particles are polydisperse with $r_p = 0.2$ and if not specified otherwise, it is assumed that $L=50$ with the total of $N_p\approx 2000$ particles.  Larger domain simulations 
are carried out with  $N_p$ up to $\approx 10, 000$.   Since the focus of the present work is on exploring universality of the force network, we have not 
made any particular effort to match the parameters between the simulations and the experiments discussed in Sec.~\ref{sec:exp}.

\begin{figure}[t!]
 \centering
 \includegraphics[width=2.1in]{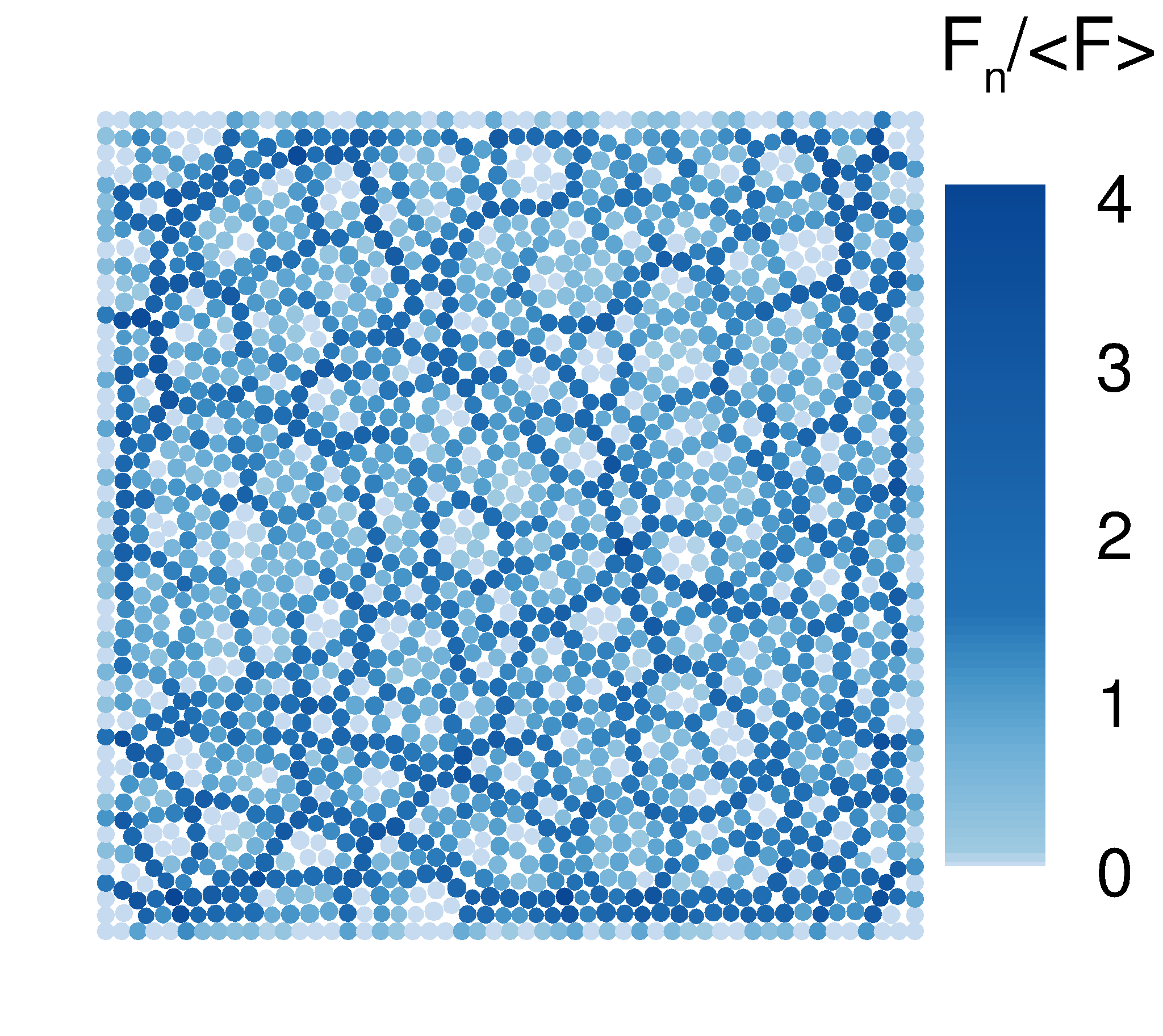}
 \caption{ Reference system at $\rho=0.9$.}\label{num_fn}
\end{figure}

\section{Force Networks and Scaling Laws}\label{sec:results}

In earlier work~\cite{kovalcinova_15}, we discussed the percolation and jamming transitions that take place as the system is exposed
to compression. We identified the packing fractions at which these transitions occur as $\rho_c$ (percolation) and $\rho_J$ (jamming).   
For the present purposes, the most relevant finding is that for the repulsive systems, $\rho_c$ and $\rho_J$ are very close and, 
in the limit of quasi-static compression, the two transitions coincide and $\rho_c = \rho_J$.
In this paper we focus on the systems such that $\rho>\rho_c$, and in particular on the properties of the force networks. 
Figure~\ref{num_fn} shows an example of a compressed packing, with the particles  color coded according to the total normal force, $F_n$, normalized by 
the average normal force, $<F>$ (we will focus only on the normal forces in the present work).  The properties of these networks depend on the force threshold,
$\bar F$, such that only the particles with $\bar F \le (F_n/<F>)$ are included.   We will now proceed to use the tools of  percolation theory to 
study cluster size distribution and mean cluster size as $\bar F$ varies, considering force networks to be
composed of the particles and inter-particle forces that can be thought of as nodes and bonds, consecutively.
 
We start by introducing the cluster number, $n_s$, representing the average (over all realizations) number of clusters with $s$ particles; note that $n_s$ depends on 
$\rho$ and $\bar F$.  From the percolation theory~\cite{stauffer}, we know that $n_s$ at the percolation force threshold, $f_p$, can be characterized by the 
following scaling law 
\begin{equation}
n_s \propto s^{-\tau},\label{scaling_ns}
\end{equation} 
with the Fisher exponent $\tau$.  The percolation force threshold, $f_p$, is
defined here as the one for which percolation probability is larger than $0.5$,
as in~\cite{kovalcinova_15}. 

Using $n_s$, we can define the mean cluster size, $S(\bar F)$, as
\begin{equation}
S(\bar F) = \frac{\sum_s^\prime s^2 n_s}{\sum_s^\prime s n_s}
\label{S_equation}
\end{equation}
where $\sum^\prime_s$ denotes the sum over the non-percolating clusters of size $s$. 

The scaling law for the mean cluster size, $S(\bar F)$, is according to~\cite{stauffer} given by 
\begin{equation}
S(\bar F) = A\, N^{\phi} \mathcal M_2 \left( B\, (\bar F - f_c) N^{1\over2\nu} \right),
\label{eq_S_scaling}
\end{equation}
where $A,B$ are the coefficients independent of the system size, $\phi,\nu$ are two critical exponents with 
$\phi=(3-\tau)/(\tau-1)$ and $f_c$ is a critical force threshold found from collapse of rescaled $S$ curves as 
described later. Note that $f_p$ and $f_c$ do not necessarily agree; we will discuss this issue later in the text. 
Here, $N$ is the total number of contacts in the system, (excluding the contacts with the wall particles), and 
$\mathcal M_2(\cdot)$ is the second moment of the probability distribution of cluster size $s$.
The question is whether there is an universal set of parameters $\phi,~\nu$ such that the $S(\bar F)N^{-\phi}$ 
curves obtained for different systems, collapse onto a single curve.  

\subsection{Computing the scaling parameters} \label{sec:comp}

To find the exponents $\phi, \nu$ and the parameter $f_c$, we follow the procedure similar to the one described in~\cite{ostojic}.
For a given simulation, we find the cluster number, $n_s$, for cluster size, $s$, ranging from $s=1$ up to $s=N_p$, given a force threshold $\bar F$. The cluster search is 
performed over the range $\bar F \in [0,5]$ with $501$ discrete levels. Then, the mean cluster size, $S(\bar F)$, is computed using Eq.~(\ref{S_equation}). 
The computation of $n_s$ and $S(\bar F)$ is performed for the systems characterized by (wall length) $L=25,~50,~75$, and $100$, and for the discrete set of $\rho$'s, such that $\rho>\rho_c$.   

\begin{figure}[t!]
\centering
{\includegraphics[width=1.65in, height=1.225in]{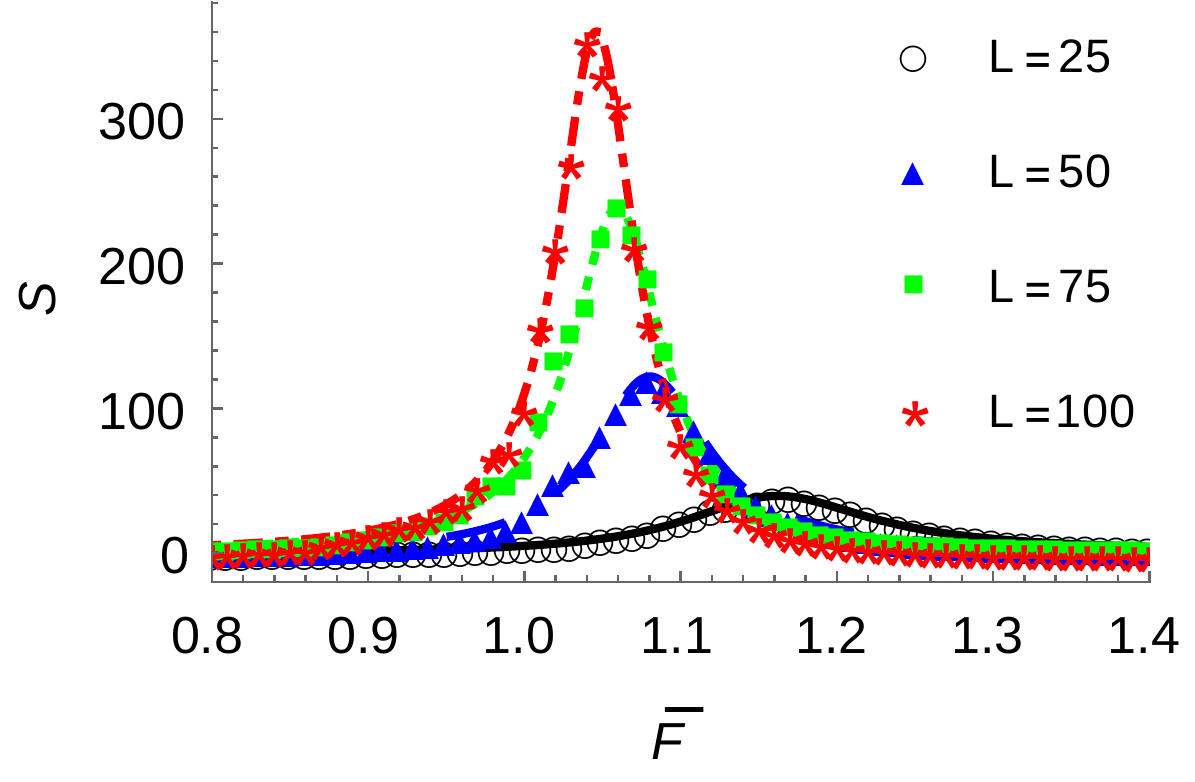}}
{\includegraphics[width=1.65in, height=1.225in]{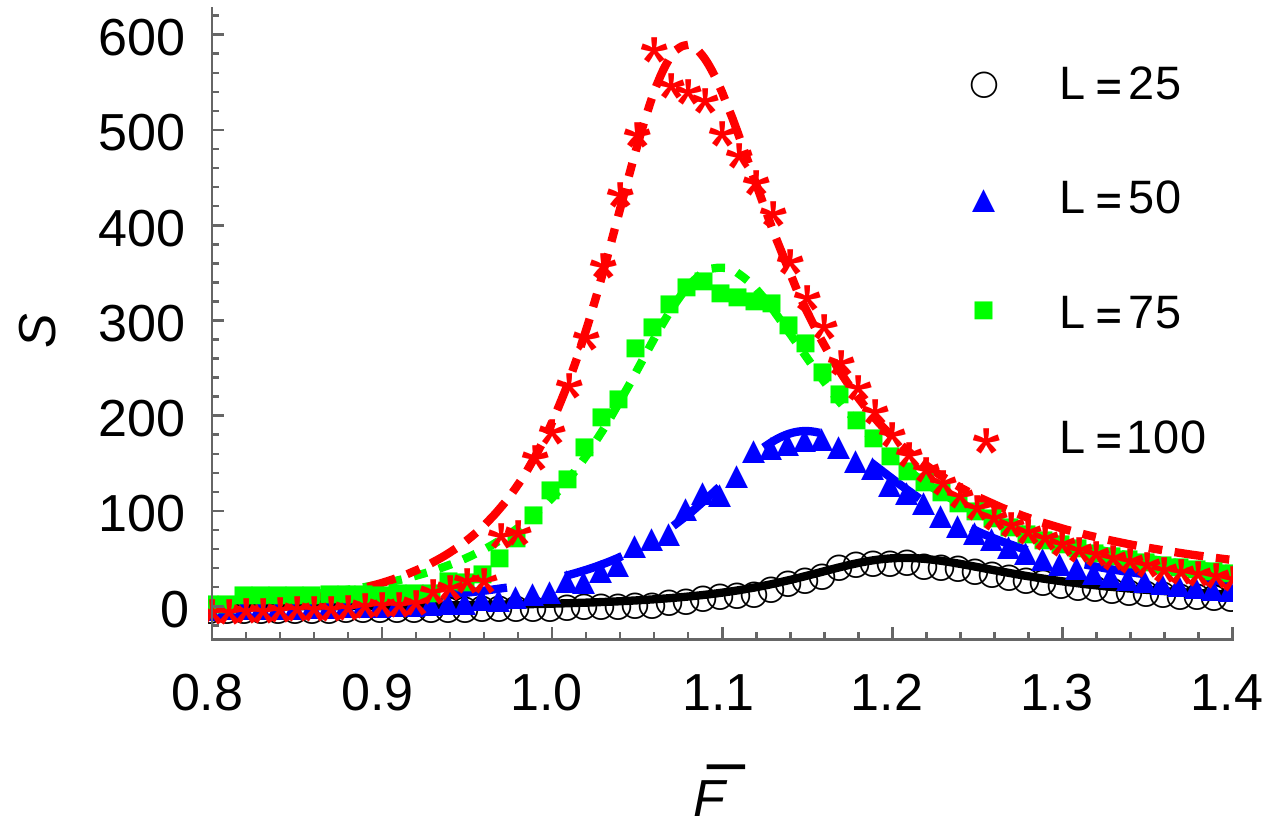}}\\
{\includegraphics[width=1.65in]{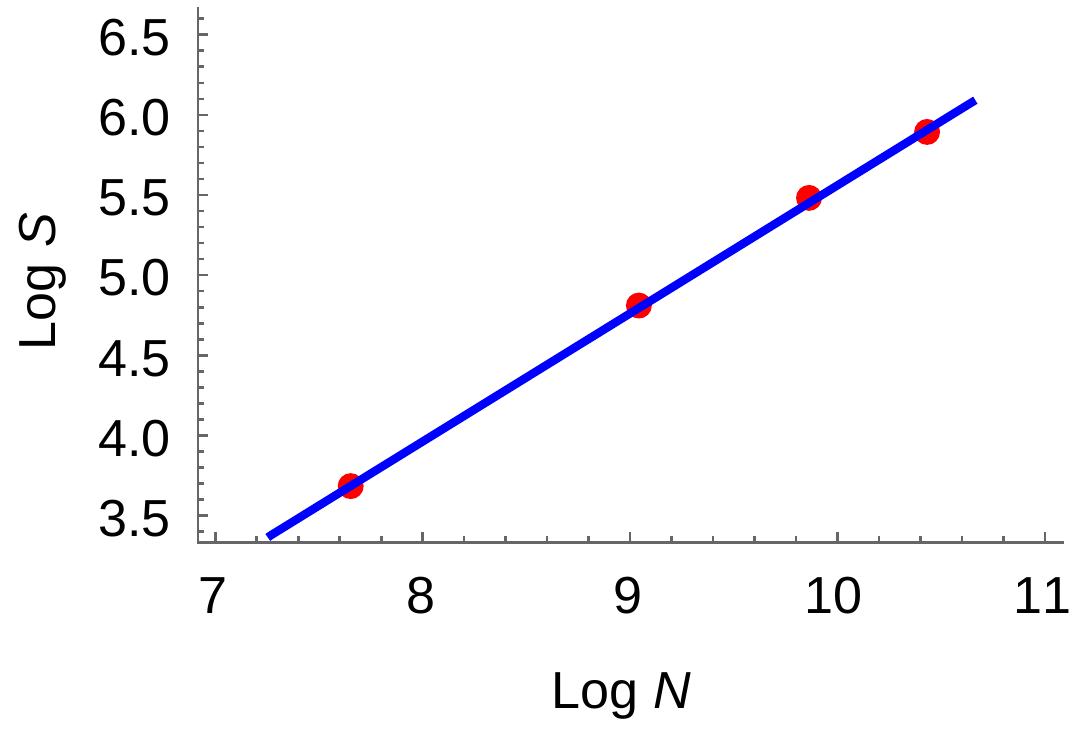}}
{\includegraphics[width=1.65in]{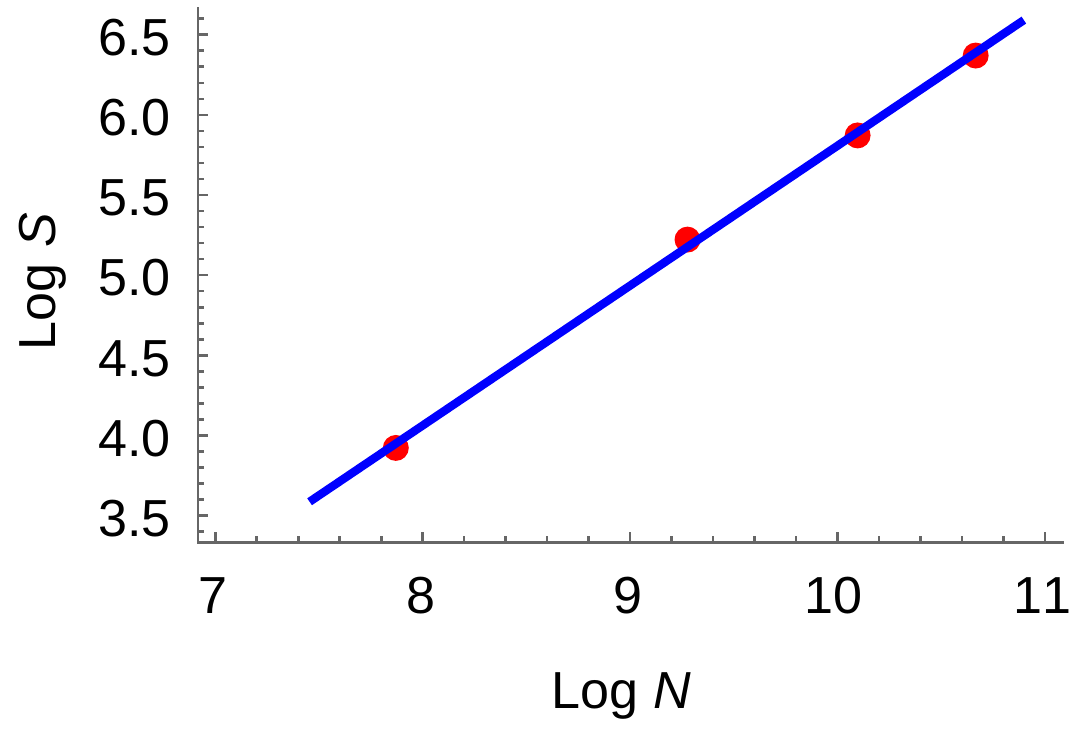}}\\
\subfloat[]{\includegraphics[width=1.75in, height=1.25in]{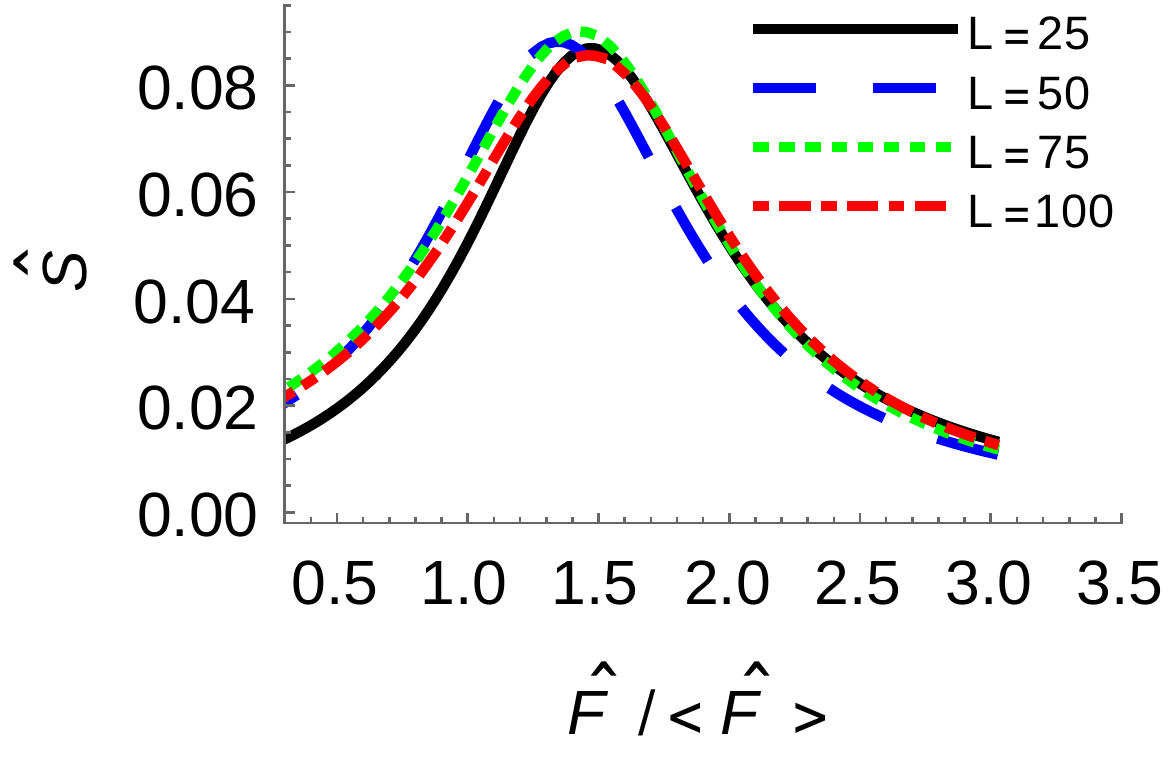}}
\subfloat[]{\includegraphics[width=1.75in, height=1.25in]{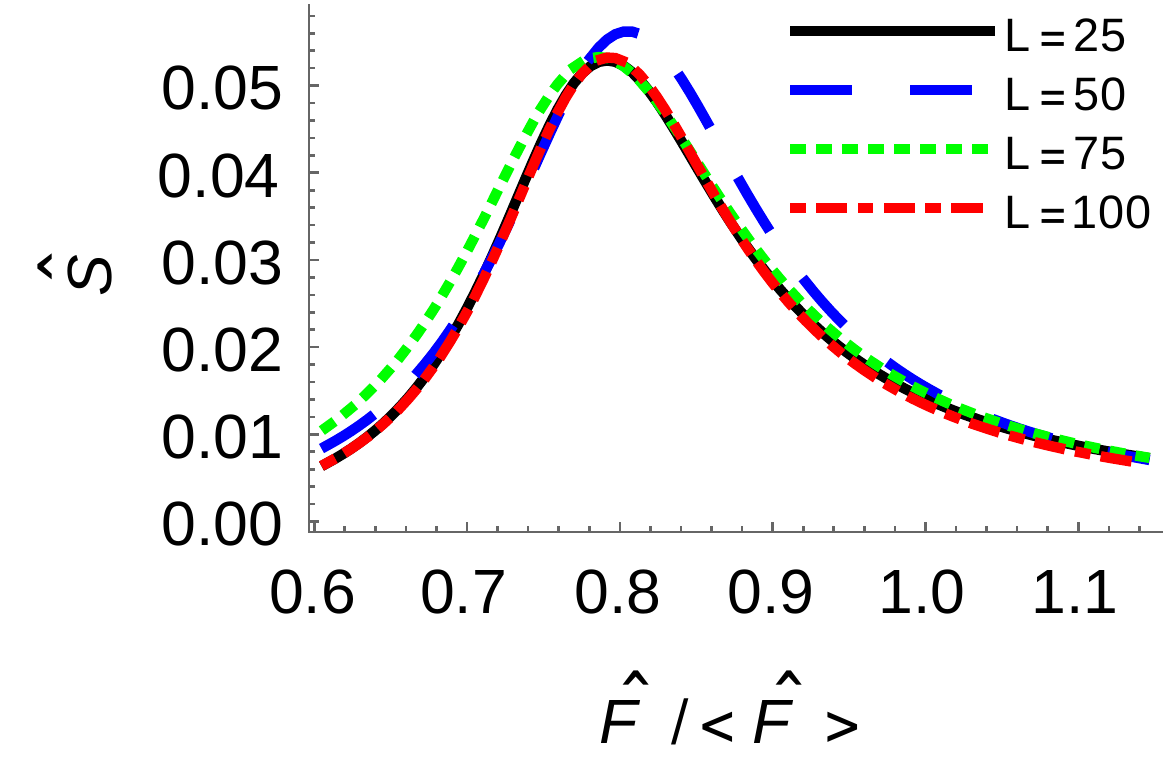}}
\caption{ (a) Reference and (b) $r_p=0.0, \mu=0.0$ system at $\rho=0.9$ showing (top to bottom) mean cluster size $S$; magnitude of the peak of $S$ versus the total number of contacts, $N$, for different 
system sizes, $L$; and collapse of the rescaled curves, $\hat S$ vs. force threshold, $\hat F$, normalized by the rescaled average force, $ <\hat F>$
(see the text for definitions of the rescaled quantities).   Note different range of axes in (a) and (b).}\label{S_allsize}
\end{figure}

Figure~\ref{S_allsize}(a) (top) shows an example of our results for $S(\bar F)$ for various $L$'s for the reference system at $\rho = 0.9$, averaged over
$120$ realizations.   As expected, the magnitude of the peak of  $S(\bar F)$ is an increasing function of $L$: according to
percolation theory~\cite{stauffer}, $S\propto L^{\beta}$, $\beta>0$ at the percolation threshold.    We note 
from Fig.~\ref{S_allsize}(a) (top) that $f_p$ (corresponding to the peak of the $S(\bar F)$ curves~\cite{stauffer})
is a decreasing function of $L$. For all systems and system size considered 
in the present work, the values of $f_p$ are
in the range $ [1.05,1.25]$.

From the magnitude of the peaks of $S(\bar F)$, one can determine the optimal critical exponent $\phi$; for $\mathcal M_2$ to be a ``universal'' curve regardless of the 
system size, $N^{-\phi}$ and $S(\bar F)$ have to balance each other as $L$ varies. The exponent $\phi$ is obtained from the linear regression through the peaks of 
$\log S(\bar F)$ as a function of $\log N$. Figure~\ref{S_allsize} (middle) shows the values of peaks as a function of $N$ in a log-log scale and a fit leading to a value of $\phi=0.80\pm 0.03$ within the $95\%$ confidence interval. 

Using the optimal value for $\phi$, the remaining two parameters, $f_c$ and $\nu$, are determined by attempting to collapse the average $S(\bar F)$ curves, 
such as the ones shown in Fig.~\ref{S_allsize} (top), around the maxima. 
We define the (large) range of values of $f_c$ and  $\nu$ over which the search is carried out: $f_c\in[0.5,2.5],\nu\in[0.5,14]$ with a discretization step $10^{-2}$. The search range for both parameters is chosen in a way such that we always find optimal values of $f_c$ and $\nu$; we verified that our results do not change if we assume larger range.
For  each $L$, we find the interval of force thresholds, $\bar F\in [a_L,b_L]$, for which $S(\bar F) \geq S_{\max}/8$, where $S_{\max} = \max\{S(\bar F)\}$.
The results are not sensitive to this specific choice of $a_L$ and $b_L$.   
For each pair $f_c,\nu$ we take the common subinterval $[a',b']=\cap[a'_L,b'_L]$ where $a'_L=(a_L-f_c)N^{1/(2\nu)}, b'_L=(a_L-f_c)N^{1/(2\nu)}$ are rescaled 
endpoints of the interval $[a_L,b_L]$.
\begin{figure}[ht!]
 \centering
 \subfloat[]{\includegraphics[width=3in]{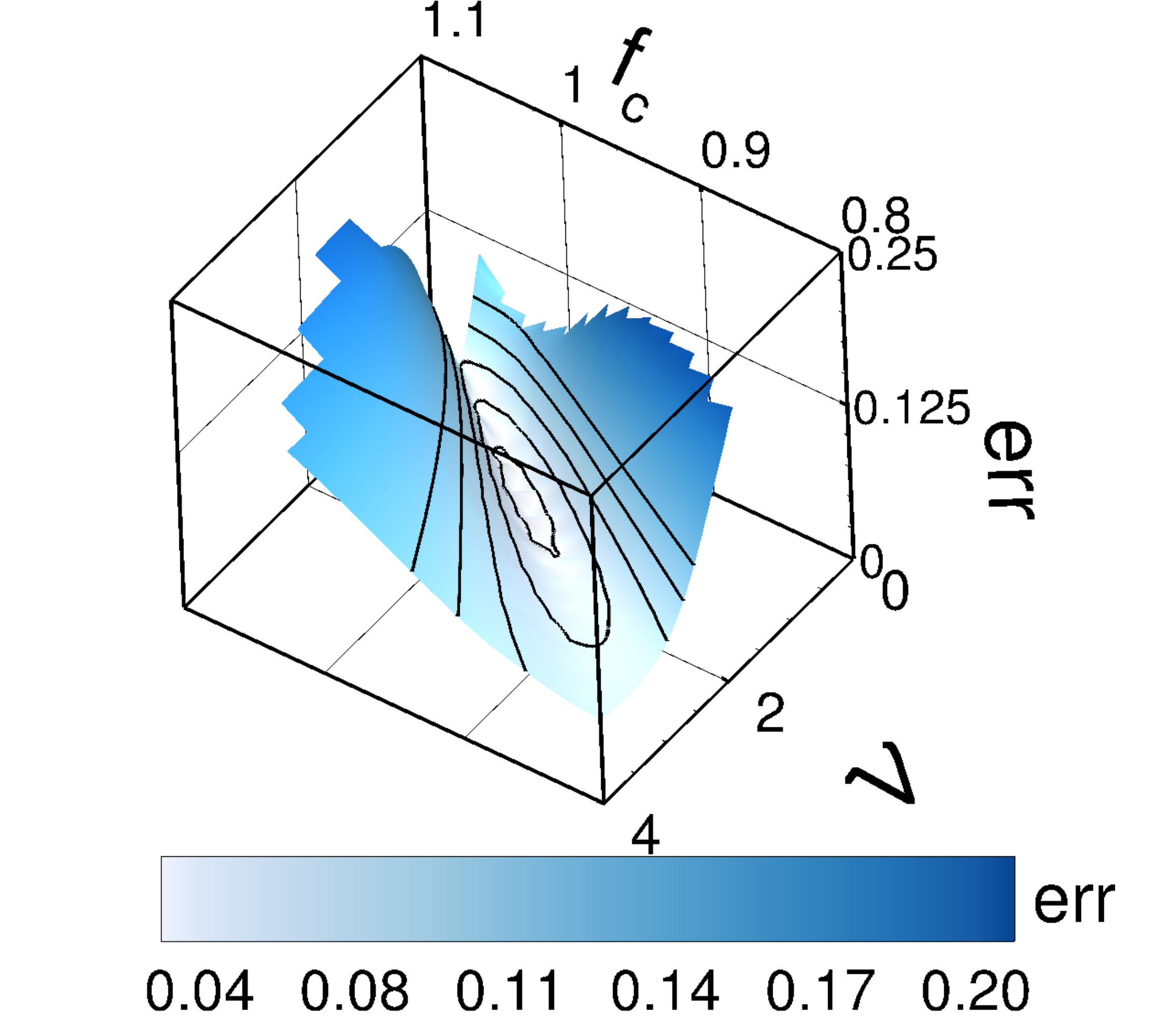}}\\
 \subfloat[]{\includegraphics[width=2.5in]{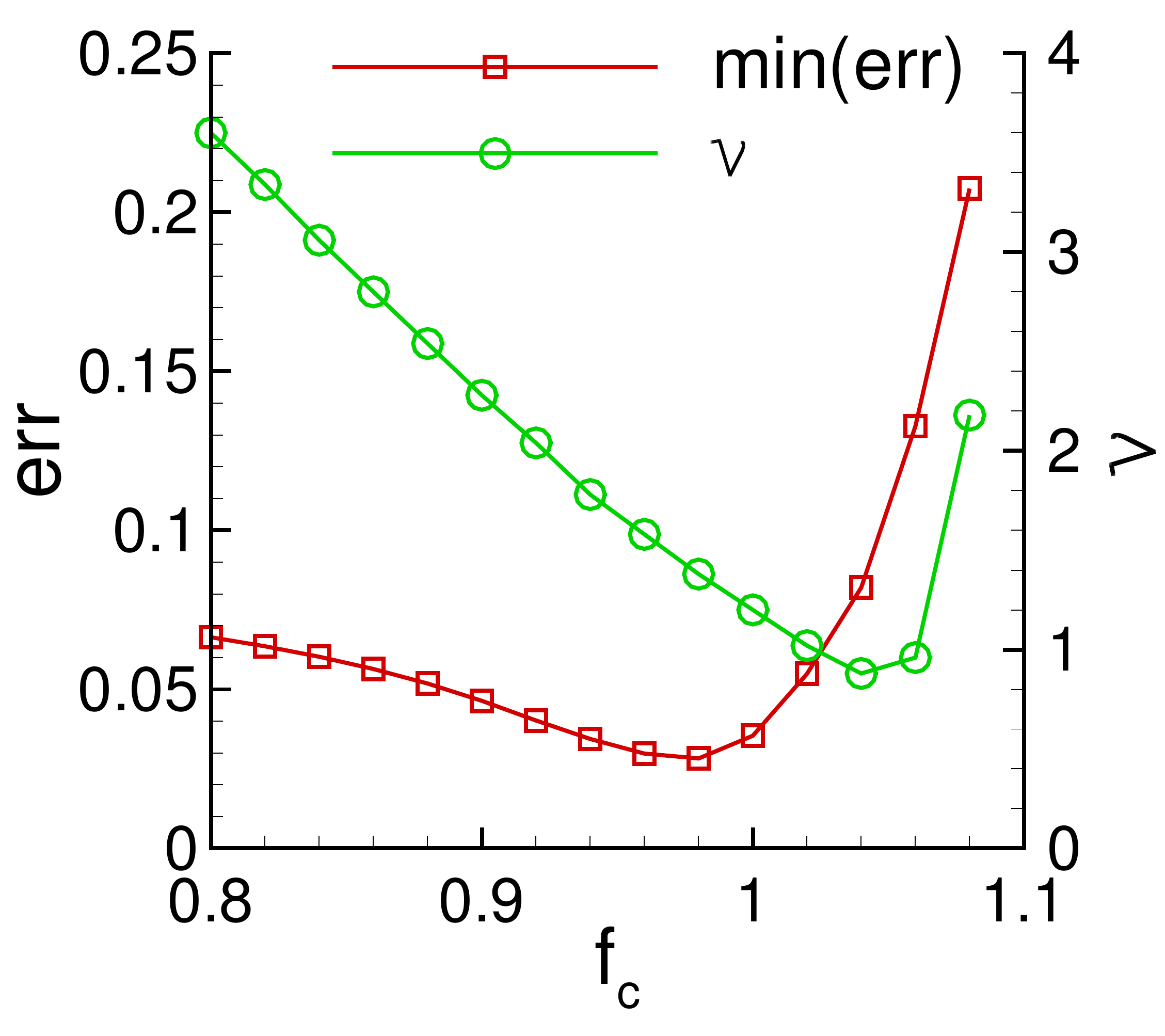}}
 \caption{
 Error plot for the reference system at $\rho =0.9$: (a) $err$ as a function of $f_c$ and $\nu$; the black lines are slices through different values of 
 $err$ and (b) minimum of $err$ (red line with squares) as a function of $f_c$; here for each $f_c$ value we choose $\nu$ that minimizes $err$; the used
values of $\nu$ are shown by the green line with circles. }\label{err_plot_r02kt08}
\end{figure}
 The optimal values of $f_c$ and $\nu$ are found by minimizing the error, $err$, defined by
\begin{eqnarray}
err&=&{1\over M}\sum_{m<n}\sum_{i=0}^{M-1}\biggl|\hat S_{L_m}(\hat F_i) - \hat S_{L_n}(\hat F_i )\biggr|~. \\
\end{eqnarray}
Here $L_m, L_n \in \{25,50,75,100\}$ are different system sizes, $\hat F_i = \bar F_i N^{-1/ (2\nu)} + f_c$ and 
$\hat S_{L_m}(\hat F_i) = S_{L_m}(\hat F_i N^{-\phi})$. We choose $\hat F_i = a'+ id_F$ with discretization step 
$ d_F = {(b'-a')/(M-1)} $  and $i = 0 \dots (M-1)$ with total of $M=100$ discretization points. 
Note that the expression for $err$ does not depend on the size of the interval over which the collapse of the curves is attempted. Figure~\ref{S_allsize} (bottom) shows the collapse of the $S$ curves as a function of the rescaled force threshold normalized by the average force threshold, $\hat F/<\hat F>$; visual inspection suggests that indeed a good collapse was found and we continue by discussing the error using the optimal values of $f_c,~\nu$. 

Figure~\ref{err_plot_r02kt08}(a) plots the contour of  $err$ as a function of $f_c$ and $\nu$ for the reference system at $\rho=0.9$.   More
precise information can be reached from  Fig.~\ref{err_plot_r02kt08}(b) that shows $err$ and $\nu$, that minimizes $err$, as a function of $f_c$.
We find that $err$ reaches a well defined minimum at $f_c \approx 0.98$ for $\nu \approx 1.38$.   

\begin{figure}[t!]
 \centering
 \subfloat[]{\includegraphics[width=1.8in]{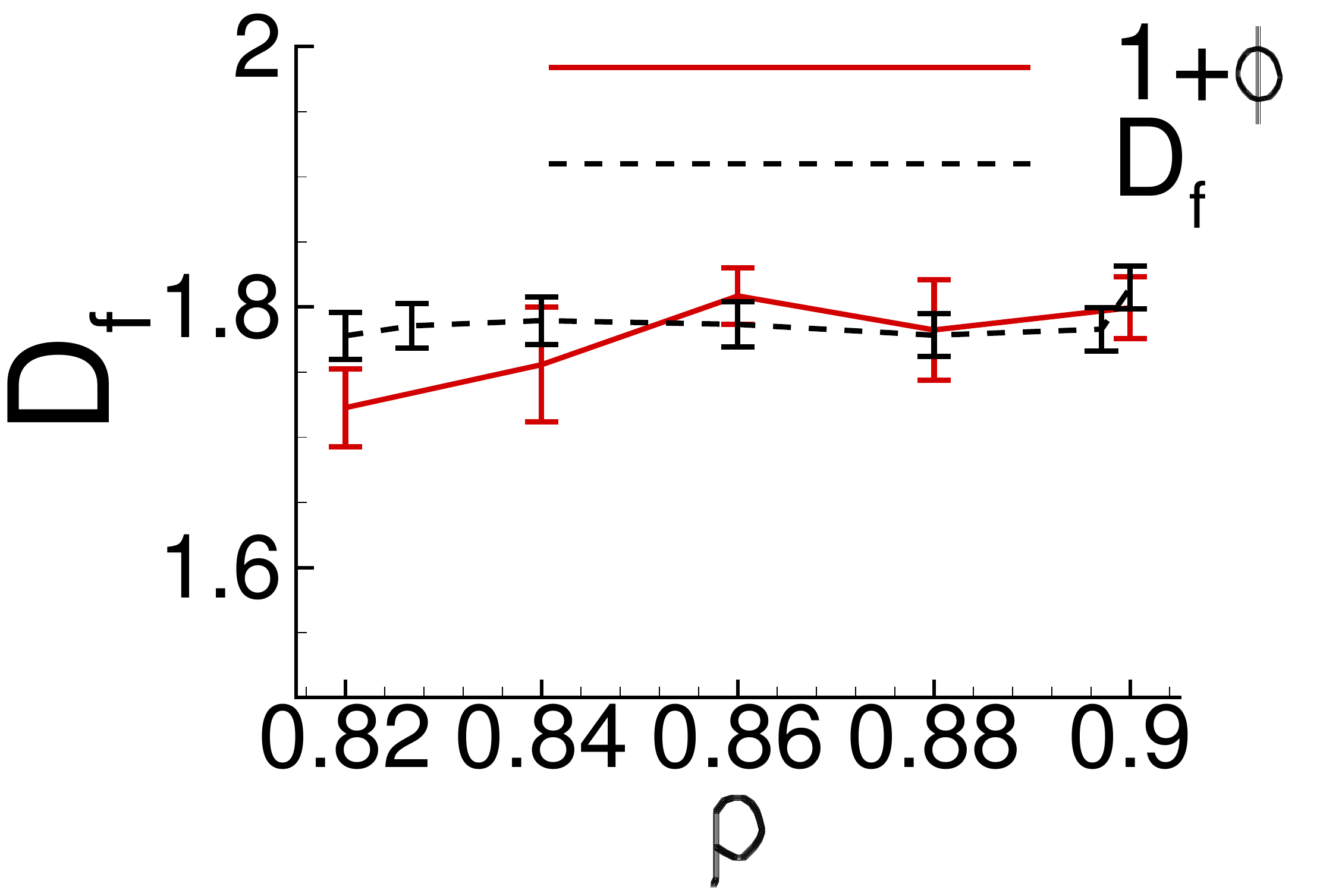}}
 \subfloat[]{\includegraphics[width=1.8in]{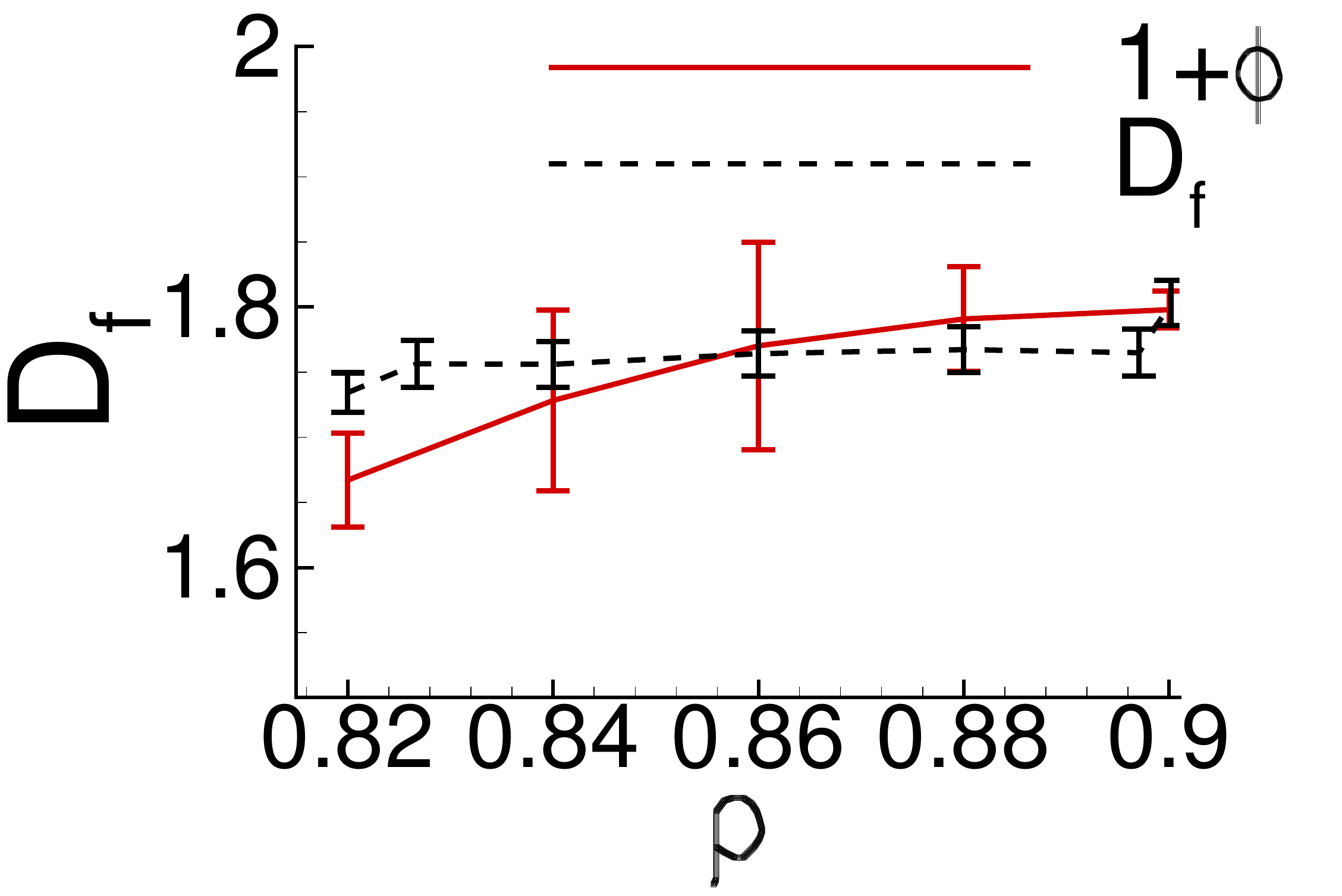}}\\
 \subfloat[]{\includegraphics[width=1.8in]{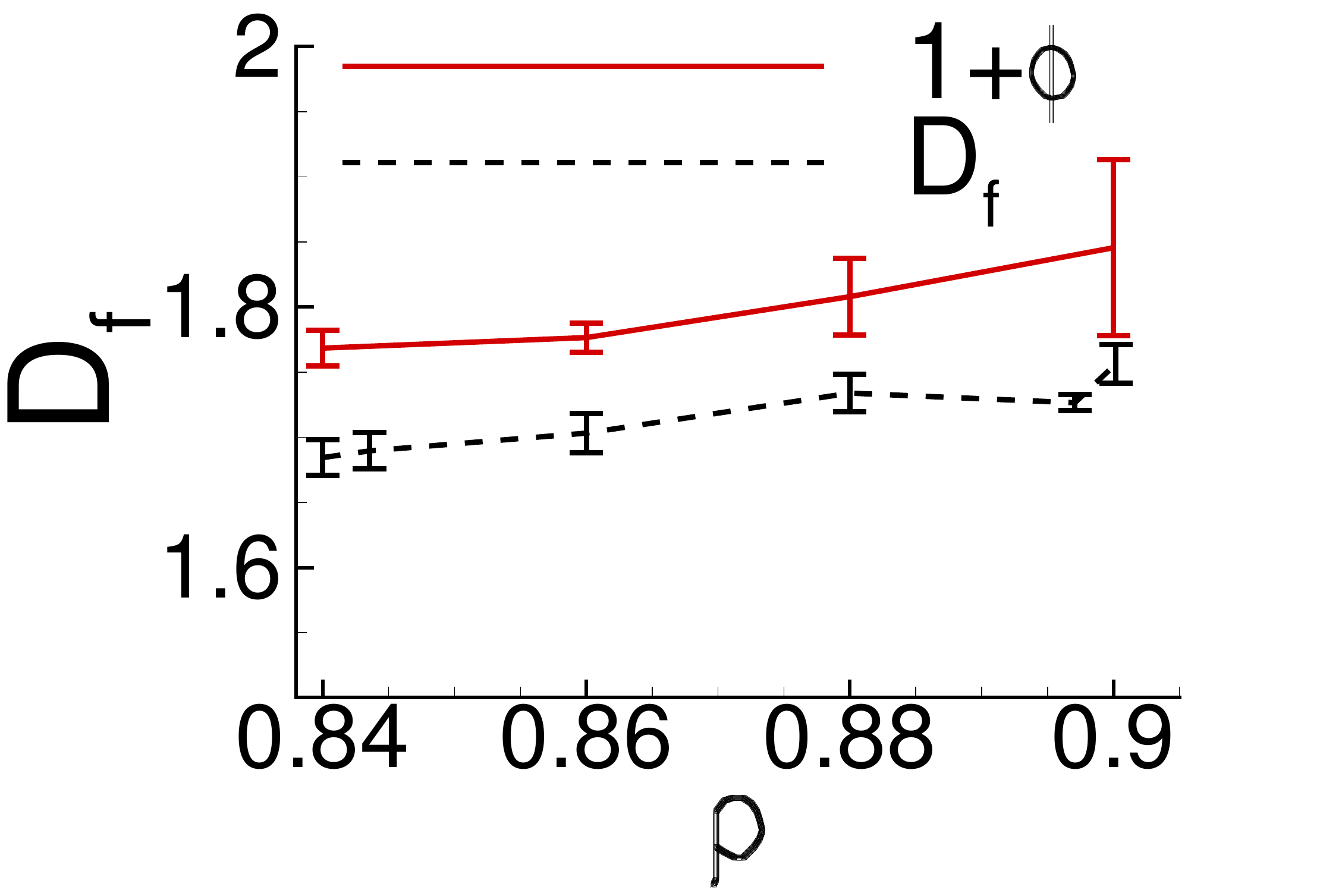}}
 \subfloat[]{\includegraphics[width=1.8in]{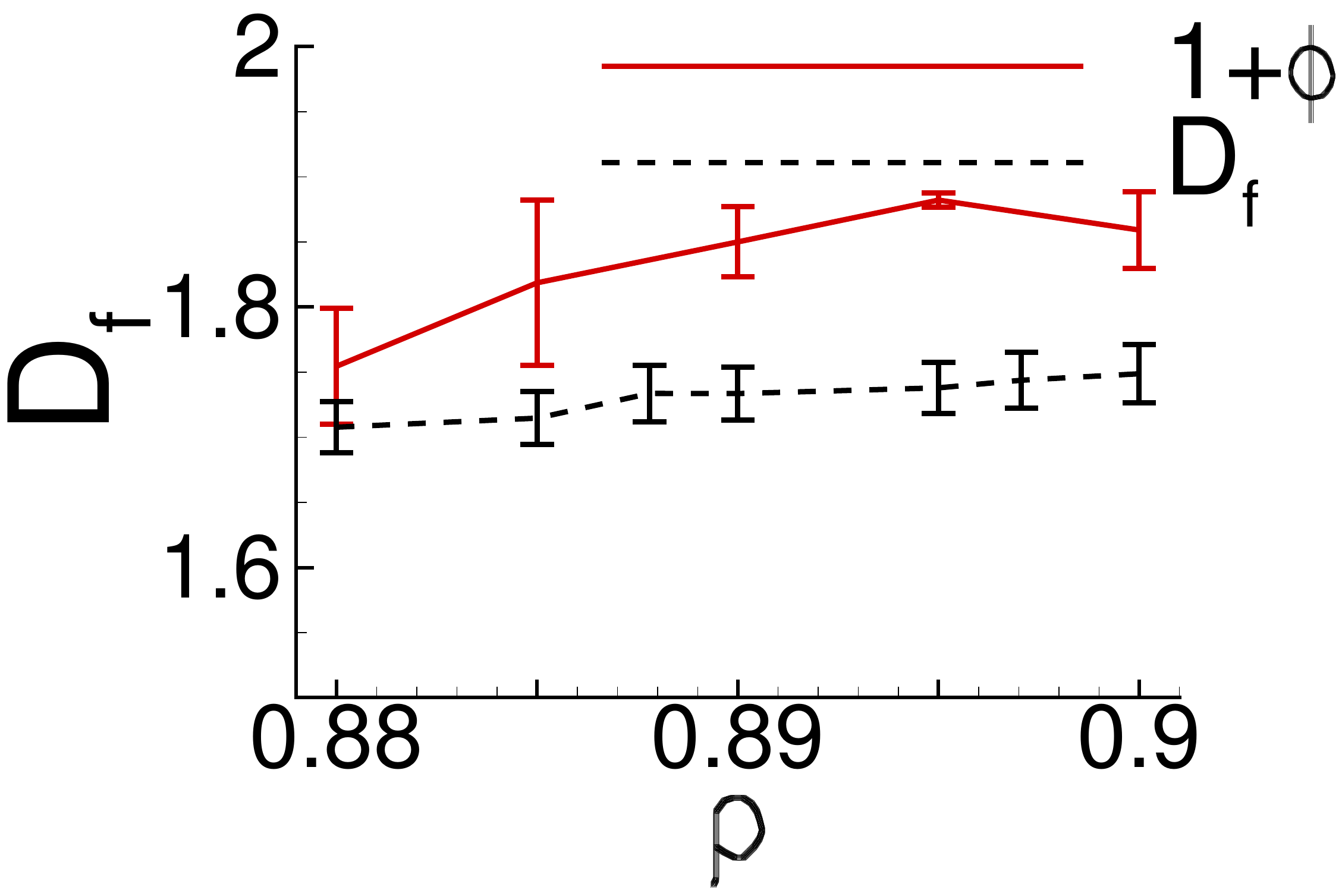}}\\ 
 \caption{Fractal dimension, $D_f$, and scaling exponent $1+\phi$ as a function of $\rho$: (a) $r_p=0.2$, $\mu=0.5$ (reference system) (b) 
 $r_p=0.0$, $\mu=0.5$ (c) $r_p=0.2$, $\mu=0.0$ and (d) $r_p=0.0$, $\mu=0.0$.  For $D_f$, the error bars represent standard error; 
 for $\phi$, the error bars are derived from the accuracy of the fit, as explained in the text. 
 The jamming packing fractions for the considered systems are $\rho_J = 0.789,~0.804,~0.827,~0.861$, respectively~\cite{kovalcinova_15}.}\label{df_vs_phi}
\end{figure}

\subsection{The scaling exponent $\phi$ and the fractal dimension}\label{sec:phi}

Before discussing how the results for $\phi,~\nu$, and $f_c$ depend on the properties of the particles, 
we mention an alternative approach to compute $\phi$. According to \cite{stauffer}, $\phi$ is related to the 
fractal dimension, $D_f$, of the percolating cluster at the percolation threshold, $f_p$, as $1+\phi = D_f$. We compute $D_f$ 
from the mass of the percolating cluster, using the Minkowski-Bouligand (or box counting) method.    For each realization 
and each $\rho$, we divide the domain into square sub-domains of the size $r$, with 
$r$ ranging from particle size up to $L$ ( $\approx 500$ discretization steps are used).  
The number of sub-domains/squares, $\mathcal N(r)$, that we need in order to cover the area occupied by 
the percolating cluster scales as 
\begin{equation}
\mathcal N(r) \sim r^{D_f}.
\label{eq_N}
\end{equation}
We compute $D_f$ for the largest system size considered, with $L=100$.    As an example, for our reference system, and for
$\rho=0.9$, we find $D_f = 1.783 \pm 0.016$.    Therefore, for the present case, we find that $D_f$ and $\phi$ are 
consistent; it is encouraging to see that the two independent procedures lead to the consistent results.  
Note also that the error in $D_f$ is smaller than the one for $\phi$; this is due to the fact that  $D_f$ is based on  
the properties of the percolating cluster that typically involves large number of particles, while the calculation of $\phi$ is 
based on smaller clusters.  Therefore, the quality of data used for calculating $D_f$ is in general much better.

We note that  the values obtained for $\phi$ are significantly lower than those given  in~\cite{ostojic} (reported value $\phi\sim0.9$), 
which is outside of the confidence interval for  $\phi$ and $D_f$ computed here.
While it is difficult to comment on the source of this difference, it may have to do with the manner in which $\phi$ is 
computed in~\cite{ostojic} - only a single domain size with $\approx 10,000$ particles was used, and then this 
domain was split into subdomains, with the largest subdomains discarded.   The remaining subdomains contain 
relatively small number of particles, leading to potential inaccuracy of the results.   

Figure~\ref{df_vs_phi} shows the independently computed values for $\phi$ and $D_f$ for the systems considered, and for the packing 
fractions above jamming, $\rho> \rho_J$; note that each of the considered systems (that differ by frictional properties and polydispersity)
jams at different $\rho_J$, listed in the caption of Fig.~\ref{df_vs_phi}.    Figure~\ref{df_vs_phi}(a) shows that for the reference system, $D_f$ and $1+\phi$ are 
in general consistent for all $\rho$'s considered, with slightly larger discrepancies close to $\rho_J$.    

\begin{figure}[h!]
 \subfloat[]{\includegraphics[width=1.8in]{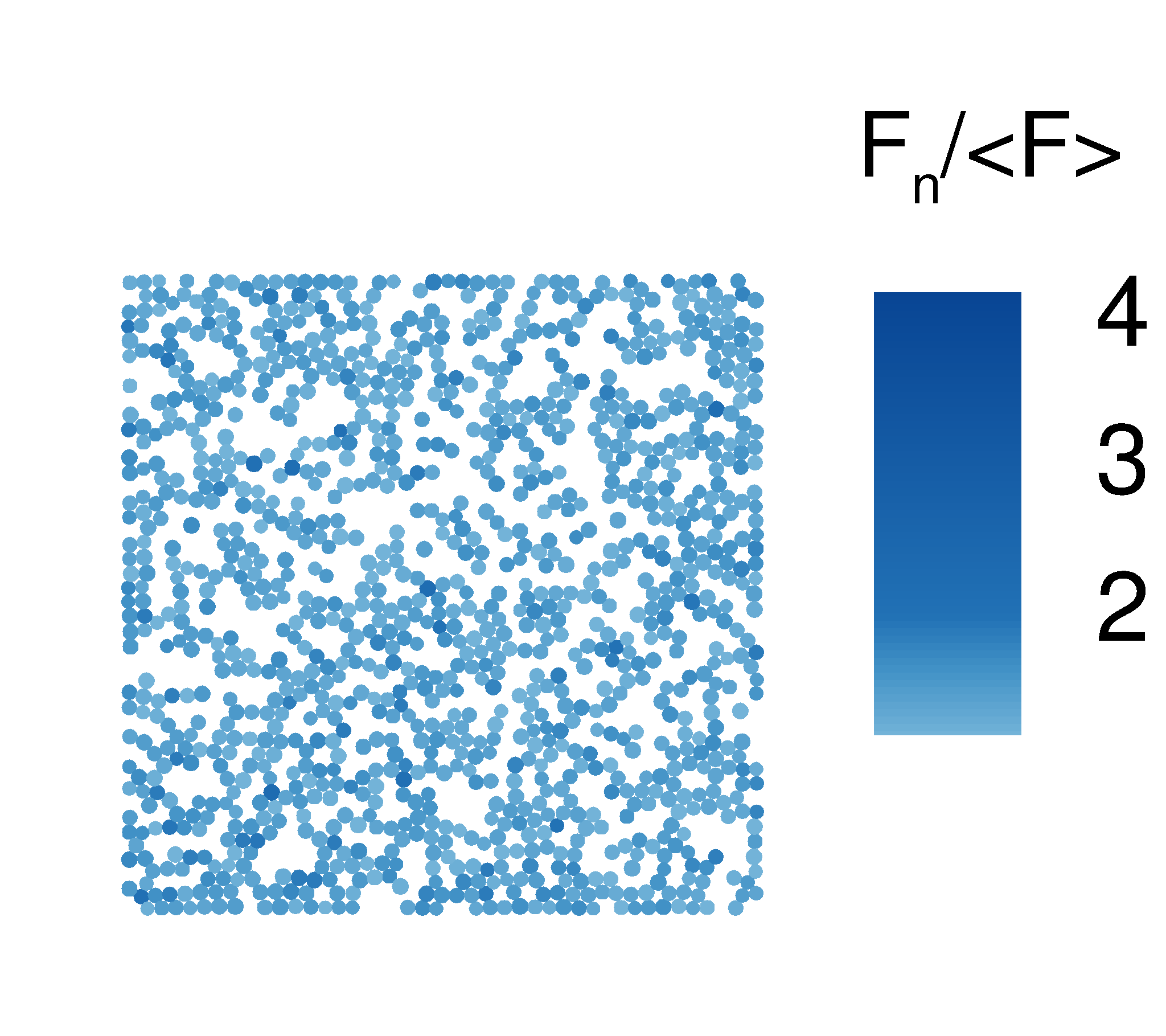}}
 \subfloat[]{\includegraphics[width=1.8in]{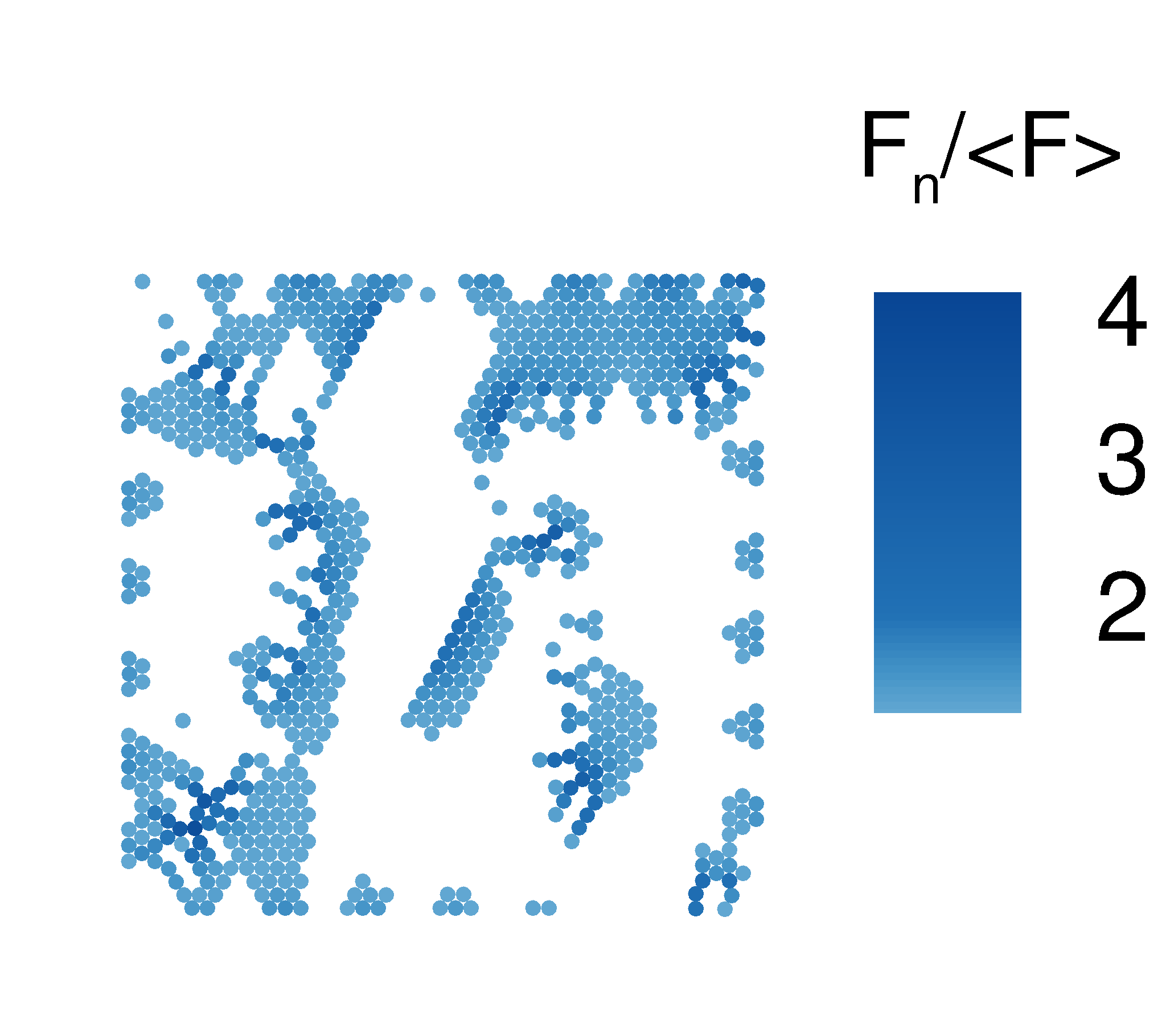}}
 \caption{(a) Reference system and (b) $r_p=0.0,\mu=0.0$ system showing only the particles with forces 
 above their respective $f_p$ at $\rho=0.9$.}\label{percolating_clusters}
\end{figure}

Figure~\ref{df_vs_phi}(b) shows the results for $D_f$ and $1+\phi$ for the $r_p=0.0, ~\mu=0.5$ system.
Similarly as for the reference case,  the values of $1+\phi$ and $D_f$ are consistent (the results for $\phi$ and $D_f$, together 
with the values of $\nu$ and $f_c$ are also given in Tables~\ref{fp_nu_friction} and~\ref{fp_nu_frictionless}).
However, for the frictionless systems, shown in 
Fig.~\ref{df_vs_phi}(c - d), we find that  there is a notable discrepancy between $1+\phi$ and $D_f$.    By comparing frictional 
and frictionless results, we note that the discrepancy comes from considerably smaller values of $D_f$ for the frictionless ones.  
This is significant, since  $D_f$ can be computed very accurately, showing clearly strong influence of friction on 
the fractal dimension.   

\subsection{Influence of particle structure on the properties of force networks}\label{sec:structure}

The obvious question is what is the source of such a large difference between frictional and frictionless systems.  The answer
appears to be a partial crystallization of the latter ones.  We discuss in more detail the structure of these systems below, but 
before that we give an illustrative example.  Figure~\ref{percolating_clusters} shows  the particles with forces above
their respective $f_p$'s for the  reference and $r_p=0.0,\mu=0.0$ systems at  $\rho = 0.9$. For the reference case we observe a large  disordered 
percolating cluster spanning the whole domain. On the other hand, the percolating cluster for the $r_p=0.0,~\mu=0.0$ system is sparse
and is built out of connected crystalline patches.   This example suggests that the  percolating cluster in partially crystallized systems
fills out much smaller portion of the domain compared to the percolating cluster in disordered packings and therefore has
much smaller $D_f$.     As a side note, we observe that for partially crystallized systems, the crystalline zones do not percolate
themselves. Instead the percolation is established by the particles at the boundaries of crystalline zones that typically sustain
larger forces than the particles that form crystals;  these highly stressed particles at the crystalline boundaries strongly influence
the topology of force networks, as discussed also in~\cite{pre13}.  In contrast to $D_f$, the values of
$\phi$ are larger for frictionless systems compared to frictional ones, showing that the expected relation between $\phi$ and $D_f$ is not
satisfied anymore.  Further discussion of the connection between $\phi$ and $D_f$ is in place; at this point we conjecture that 
the systems that crystallize are a part of a different class in terms of the scaling law of $S$.

\begin{figure}[t!]
 {\includegraphics[width=1.8in]{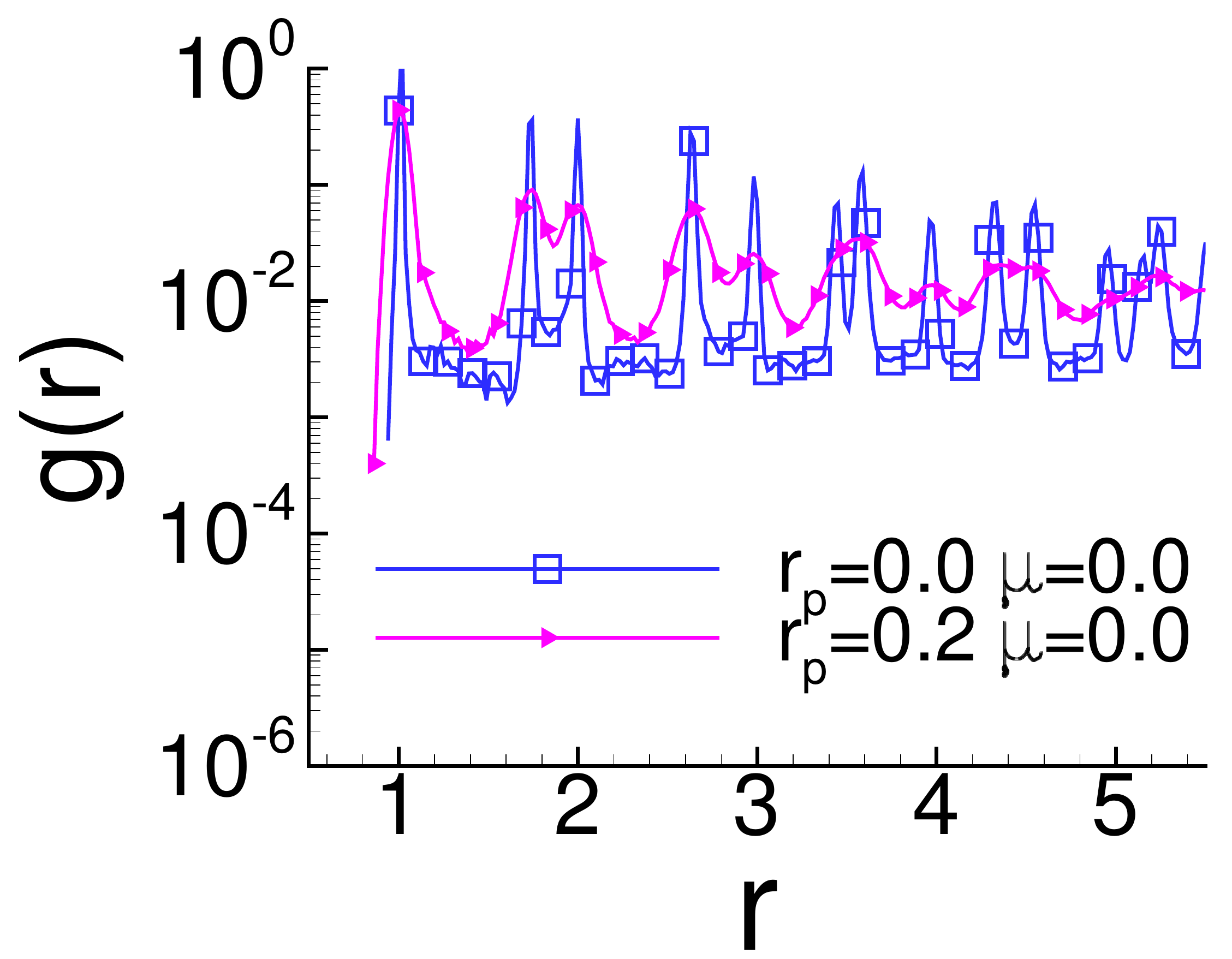}}{\includegraphics[width=1.8in]{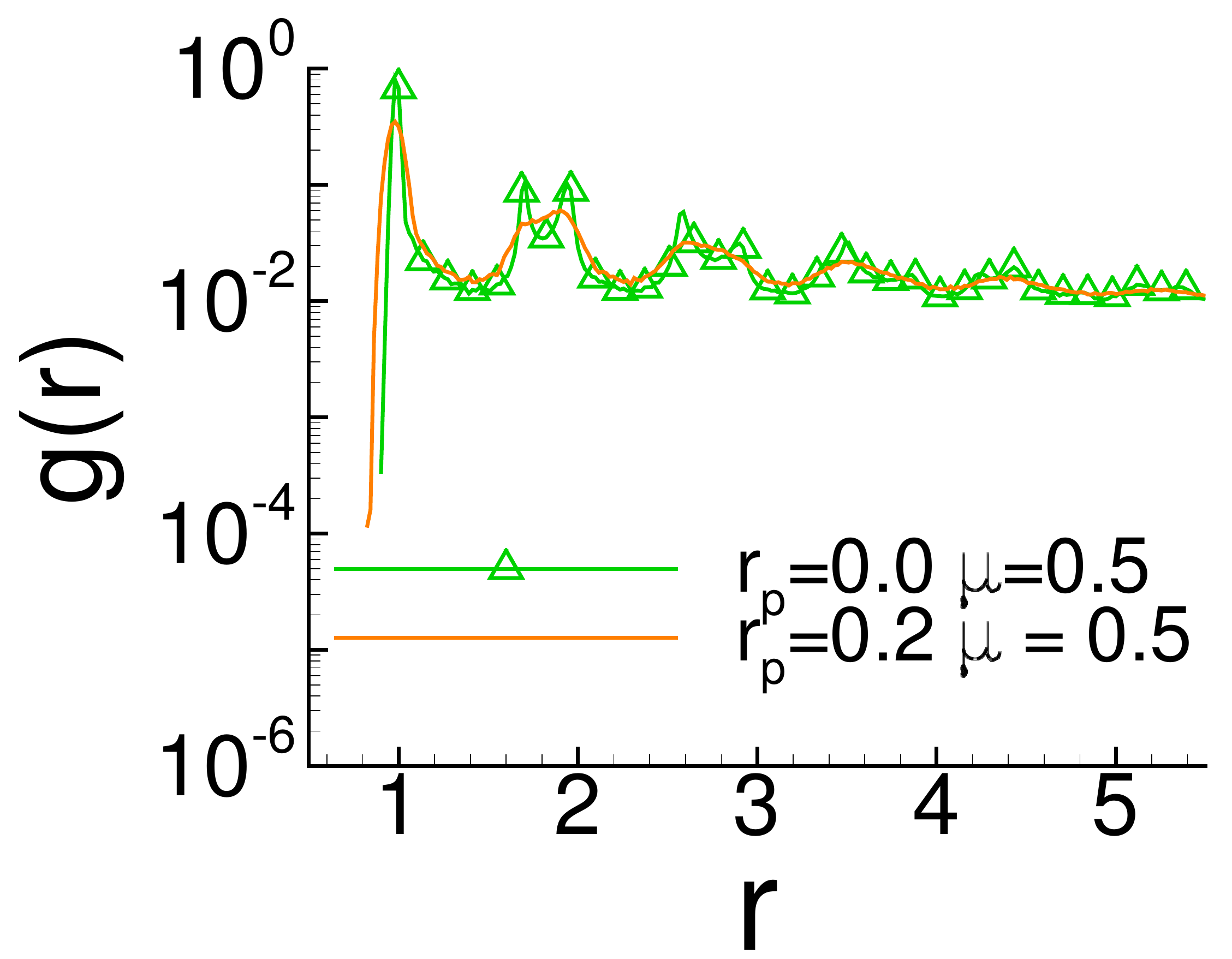}}\\
 \subfloat[]{\includegraphics[width=1.8in]{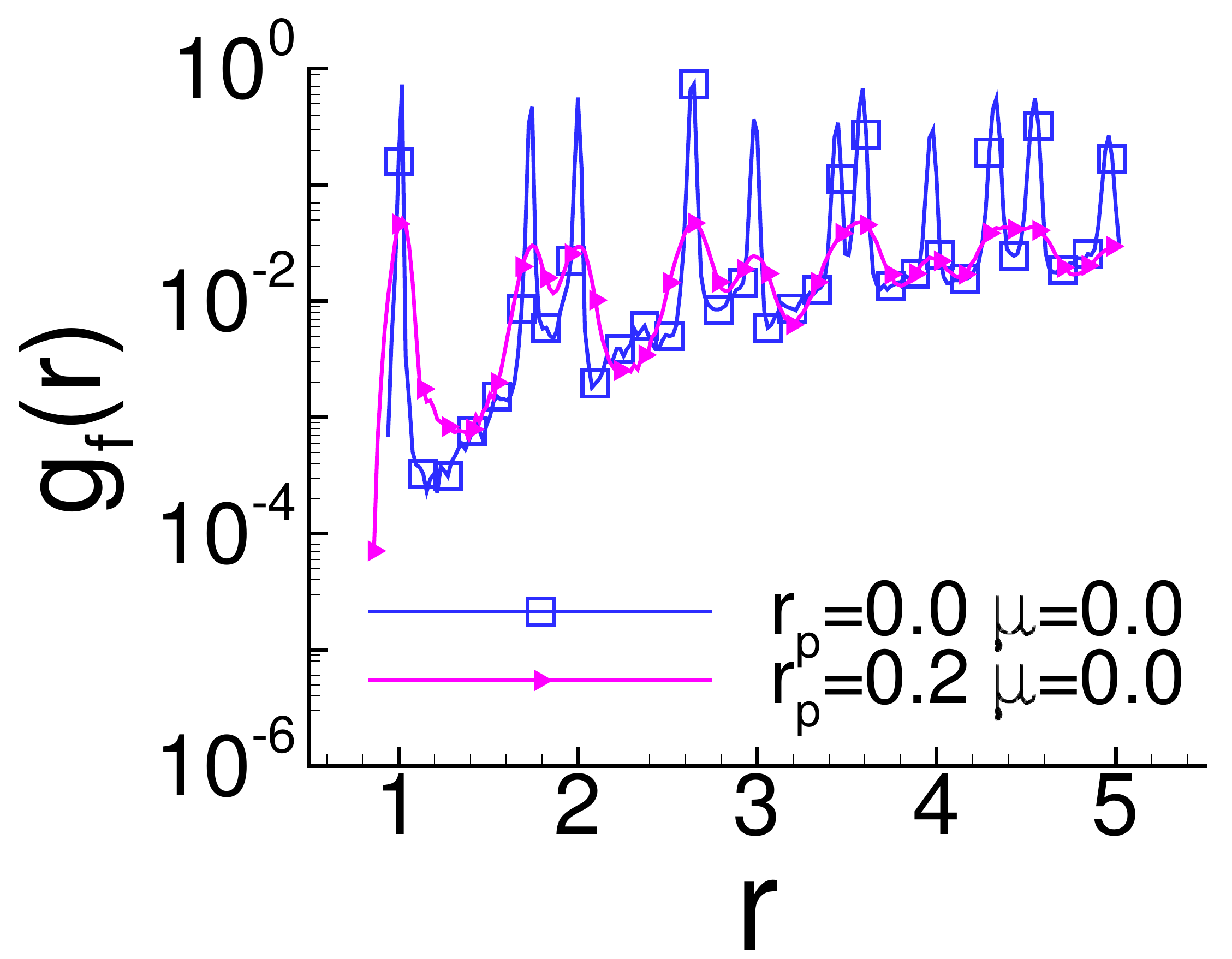}}
 \subfloat[]{\includegraphics[width=1.8in]{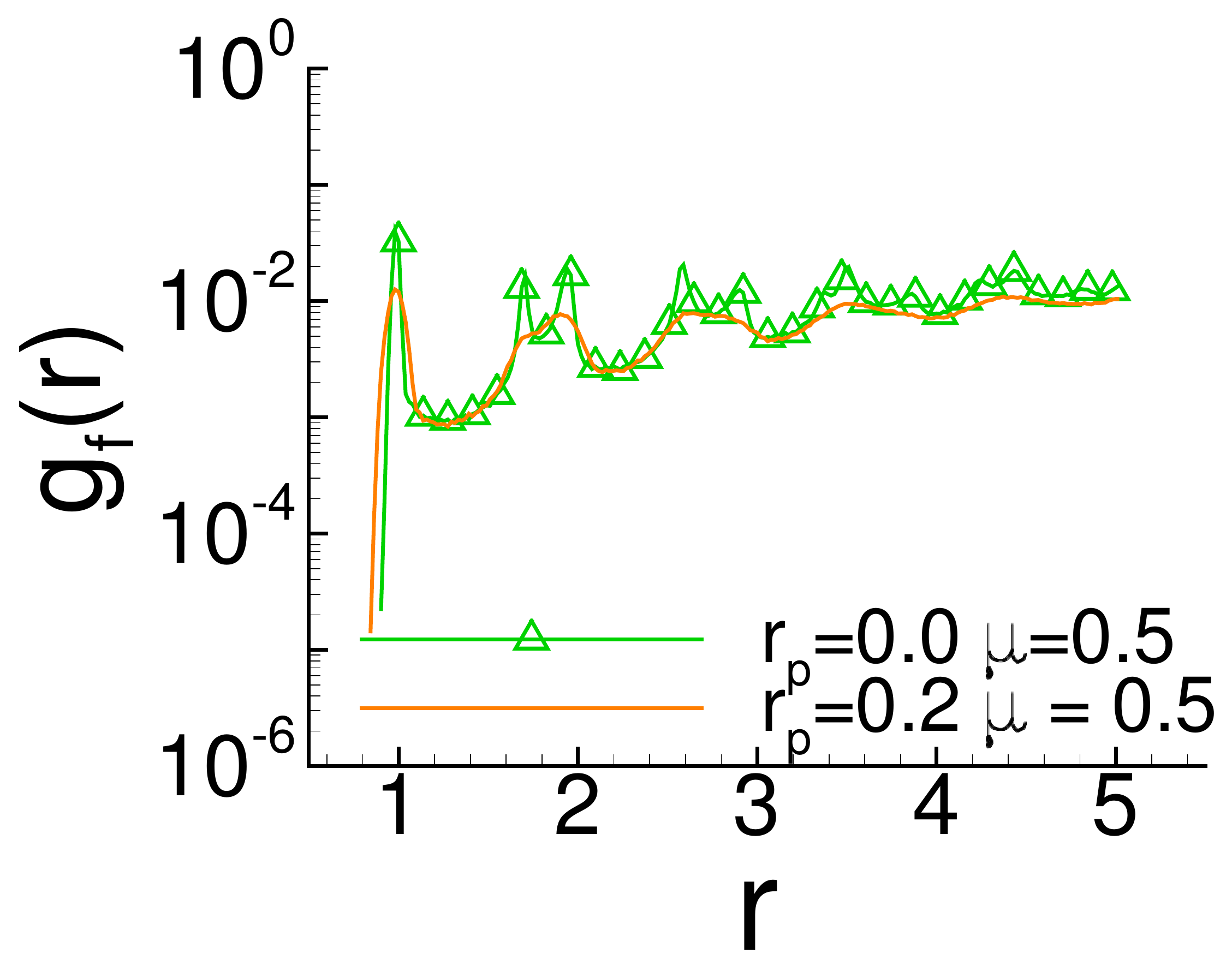}}
 \caption{Pair correlation function, $g(r)$, and force correlation function, $g_f(r)$, at $\rho=0.9$ for (a) frictionless systems,  and (b) frictional systems.}\label{correlation}
\end{figure}
\begin{figure}[ht!]
 \includegraphics[width=2.5in]{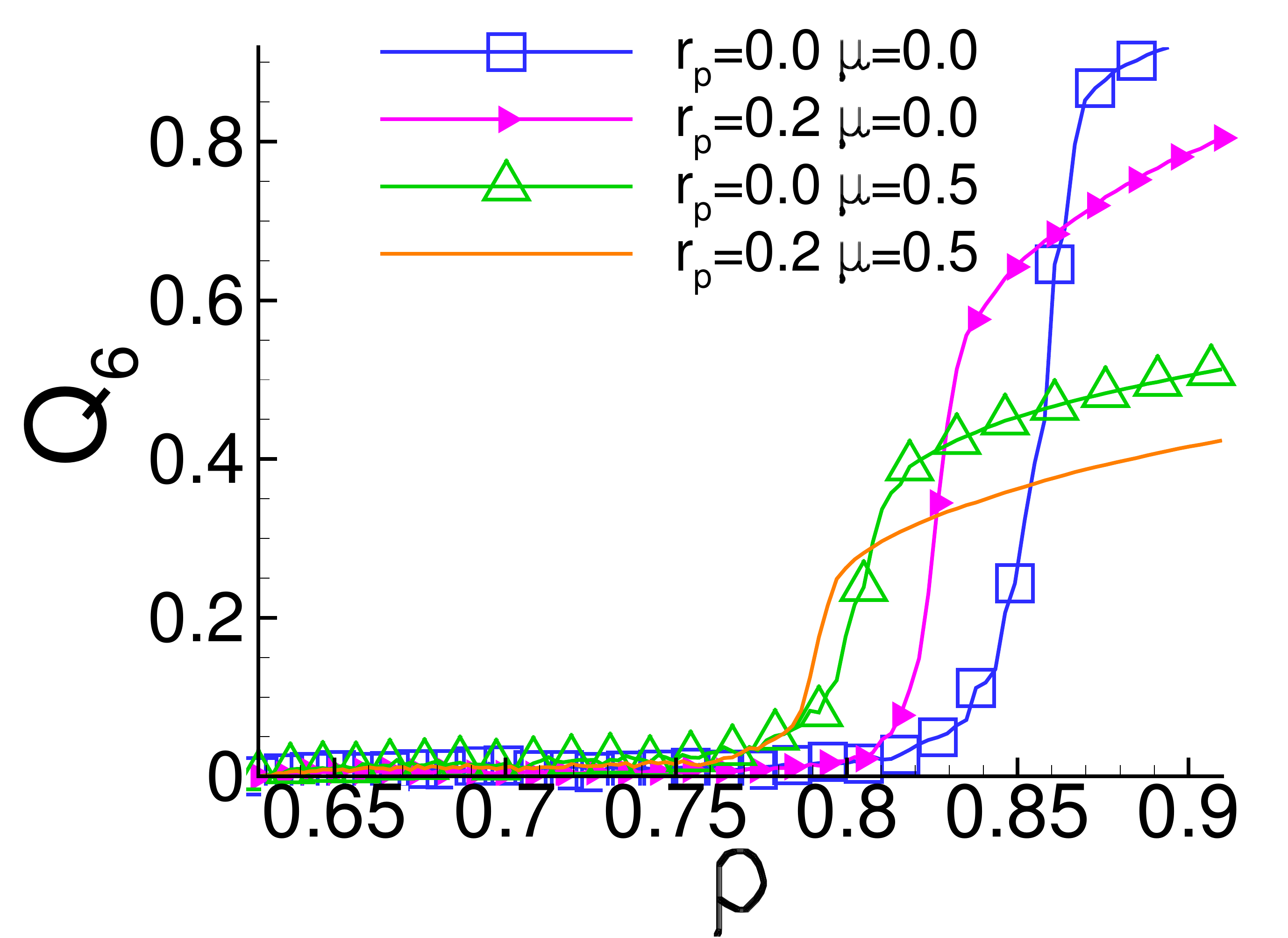}
 \caption{Order parameter showing distribution of the angles between contacts.}\label{Q6}
\end{figure}

We proceed by discussing structural properties and partial crystallization for the considered systems.
The level of crystallization is computed for the largest packing fraction, $\rho=0.9$, for all considered systems, by 
computing the pair correlation function, $g(r)$. The level of ordering of force networks is found from the force correlation function given by
\begin{align}
g_f(r) = \frac{\sum_i \sum_{j>i} \delta\left( r_{ij} - r \right)(F_i-<F>)(F_j-<F>)}{\sum_i \sum_{j>i} \delta\left( r_{ij} - r \right)}
\end{align}
where $F_i$ denotes the total normal force on $i$-th particle and $r_{ij}$ is the distance between the particles $i,j$. Figure~\ref{correlation} shows 
$g(r)$ and $g_f(r)$, averaged over $120$ realizations.   We observe a 
pronounced first peak of $g(r)$ and $g_f(r)$ for frictionless systems and a clearly split second peak, which is a sign of crystallization~\cite{torquato_98}. 
In the case of frictional systems, the first peak of $g(r)$ and $g_f(r)$ is less pronounced and clearly there is a smaller long-range correlation for both $g(r)$ and
$g_f (r)$. 

Next we discuss $Q_6$, that measures  the distribution of the angles between contacts, defined by
\begin{equation}
Q_6 = \frac{1}{N_p} \sum_i \frac{1}{C_i-1} \sum_{k=1}^{C_i-1} \cos(6 \theta_{k} ).
\end{equation}
$Q_6$ is a measure of the six-fold symmetry between contacts: here $N_p$ is the total number of particles, 
$C_i$ is the number of contacts for the $i$-th particle, and $\theta_k$ is the angle between two consecutive contacts.
Note that $Q_6$ is equal to $1$ for a perfect hexagonal crystal.
Figure~\ref{Q6} shows the results averaged over all realizations.  
For small $\rho$'s, $Q_6$ is small for all systems, but then, as the systems go through their respective jamming 
transitions, $Q_6$ grows.   For $\rho> \rho_J$, we observe that the frictionless systems, in  particular the monodisperse one,
are the most ordered, consistently with the results obtained by considering $g(r)$ and $g_f(r)$.   

\begin{figure}[t!]
 \includegraphics[height=1.7in]{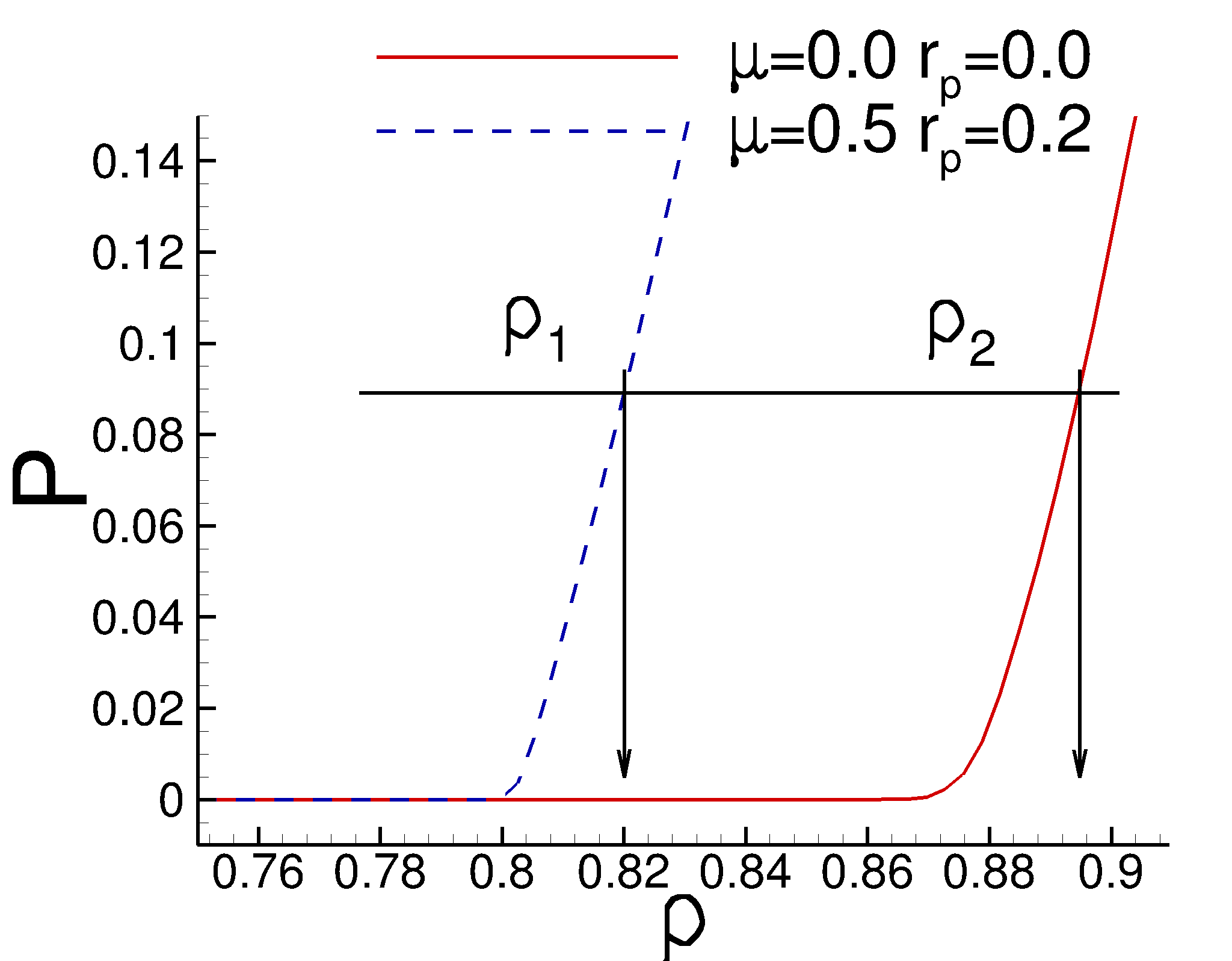}
 \caption{The pressure, $P$, on the the domain boundaries as a function of $\rho$ for the reference and $r_p=0.0, \mu=0.0$ system; $\rho_1, ~\rho_2$ correspond to the same pressure in reference and  $r_p=0.0, \mu=0.0$ system, respectively.}\label{pressure_on_walls}
\end{figure}

\subsection{Further discussion of the results for $\phi$ and the fractal dimension}\label{sec:discussion}

Here we discuss briefly two effects that could potentially influence the results presented so far: non-vanishing compression
rate, and the differences in $\rho_J$ for the systems considered.

In~\cite{kovalcinova_15}, we showed that the compression speed influences percolation and jamming 
transitions, so it is appropriate to ask whether our scaling results are influenced by non-vanishing compression
rate.  For this reason, we also consider relaxed systems, where we stop the compression and relax the particles'
velocities every $\Delta \rho = 0.02$, following the same protocol as presented in~\cite{kovalcinova_15}.   
We then compute  $1+\phi$ and $D_f$ using the same approach as discussed so far, and find that the values are 
consistent with the ones presented (figures not given for brevity).   This finding is not surprising since 
here we focus on the systems above their jamming transitions.  For such $\rho$'s, consistently with the results given in~\cite{kovalcinova_15}, 
there does not seem to be any rate dependence of the results, at least for the slow compression considered here.

We further examine whether the inconsistency of the results for $\phi$ and $D_f$ for the frictionless systems 
might arise from the proximity to the jamming transition.   As noted above, $\rho_J$ differs significantly between
the considered systems, and it reaches particularly large values for the frictionless ones.   Since we are comparing
different systems, we need to confirm that they are all in the same regime, so sufficiently far away from $\rho_J$.  
As a measure, we consider here the (dimensionless) pressure, $P$ (computed as the average force per length) on the domain boundaries.
For the sake of brevity, we focus on two representative systems here, the reference one, and the $r_p=0.0,\mu=0.0$ system.
Figure~\ref{pressure_on_walls}  shows $P$ as a function of $\rho$ for these two systems, averaged over all realizations.  Since the reference system jams for 
much smaller $\rho$, $P$ starts growing earlier.   For definite comparison, consider a particular 
packing fraction, $\rho_1 = 0.82$ for the reference system: at this $\rho$, $P$ is non-zero, and Fig.~\ref{df_vs_phi} and
Table~\ref{fp_nu_friction} show that $\phi$ and $D_f$ are consistent. Consider now $r_p=0.0,\mu=0.0$ system at the 
packing fraction $\rho_2 = 0.895$ that corresponds to the same pressure. At this $\rho$, Fig.~\ref{df_vs_phi}(d) shows inconsistent 
values of $\phi$ and $D_f$.  We conclude that the difference in the results obtained from two different methods - scaling versus fractal dimension -  
for frictionless monodisperse system does not arise from the proximity to a jamming transition.

\subsection{Continuation of the discussion of scaling parameters}\label{sec:other}

We continue with the discussion of remaining scaling parameters, $f_c$ and $\nu$, found by minimizing $err$ (the distance between the
$\hat S$ curves) for different $L$'s.  Table~\ref{fp_nu_friction} shows $f_c$ and 
$\nu$ as a function of $\rho$ for the considered frictional systems. 
While the value of $f_c$ is almost constant, $\nu$ shows the same 
decreasing trend with increasing $\rho$ for both considered frictional systems. We note, however, that the results for $\nu$ are 
different for monodisperse and polydisperse system: for the reference (polydisperse) case and sufficiently 
high $\rho$, we find $\nu\approx1.5$, consistently with~\cite{ostojic}. However, $r_p=0.0,\mu=0.5$ gives 
$\nu\approx 2$.   
Note also that for $r_p = 0.0$, $\mu = 0.5$ system and for $\rho = 0.82$ rather large $err$
is found, suggesting larger inaccuracy in the (very) large optimal value of $\nu$.  

\begin{table}[th]
  \begin{tabular}{|l|c|c|c|c|r||}
  \hline
  \multicolumn{6}{|c||}{$r_p=0.2$}\\
    \hline
    $\rho$&0.82&0.84&0.86&0.88&0.90\\
    \hline
   $f_c$&1.06&1.0&0.98&0.98&0.98\\
    \hline
   $\nu$&1.68&1.7&1.54&1.48&1.38\\
    \hline
   $D_f$&1.78&1.80&1.79&1.78&1.81\\
    \hline
   $\phi$&0.72&0.75&0.80&0.78&0.80\\
    \hline
   $err$ &0.06&0.05&0.02&0.03&0.03\\
    \hline
  \end{tabular}\begin{tabular}{|l|c|c|c|c|r|}
  \hline
  \multicolumn{5}{|c|}{$r_p=0.0$}\\
    \hline
    0.82&0.84&0.86&0.88&0.90\\
    \hline
   0.76&0.96&0.96&0.96&0.96\\
    \hline
   13.94&2.72&2.12&1.98&1.84\\
    \hline
   1.73&1.77&1.78&1.77&1.80\\
    \hline
   0.67&0.73&0.77&0.79&0.80\\
    \hline    
      0.13&0.09 &0.06 & 0.04 &0.04 \\
    \hline
  \end{tabular}

  \caption{The results are shown for $D_f$, $f_c$ and scaling exponent $\nu$ for the frictional systems; the value of $err$ gives an estimate of the accuracy of the collapse. }\label{fp_nu_friction}
\end{table}

\begin{table}[th]
\centering
  \begin{tabular}{|l|c|c|c|r||}
  \hline
  \multicolumn{5}{|c||}{$r_p=0.2$}\\
    \hline
    $\rho$&0.84&0.86&0.88&0.90\\
    \hline
   $f_c$&1.18&1.1&1.08&1.06\\
    \hline
    $\nu$&1.58&1.44&1.28&1.22\\
    \hline
   $D_f$&1.68&1.70&1.73&1.76\\
    \hline
   $\phi$&0.77&0.78&0.80&0.84\\
    \hline
   $err$ &0.01&0.01&0.08&0.01 \\
    \hline
  \end{tabular}\begin{tabular}{|c|c|c|c|c|}
  \hline
  \multicolumn{5}{|c|}{$r_p=0.0$}\\
  \hline
    0.88&0.885&0.89&0.895&0.90\\
    \hline
   0.22&0.56&0.60&0.66&0.54\\
    \hline
    13.96&8.76&6.84&5.48&6.52\\
    \hline
   1.70&1.71&1.73&1.74&1.75\\
    \hline
    0.81&0.82&0.86&0.89&0.87\\
    \hline
    0.02&0.02&0.01&0.01&0.01\\
    \hline
  \end{tabular}

  \caption{The results are shown for $D_f$, $f_c$ and scaling exponent $\nu$ for the frictionless systems; the value of $err$ gives an estimate of the accuracy of the collapse.
}\label{fp_nu_frictionless}
\end{table}

\begin{figure}[ht!]
 \centering
 \subfloat[]{\includegraphics[width=3in]{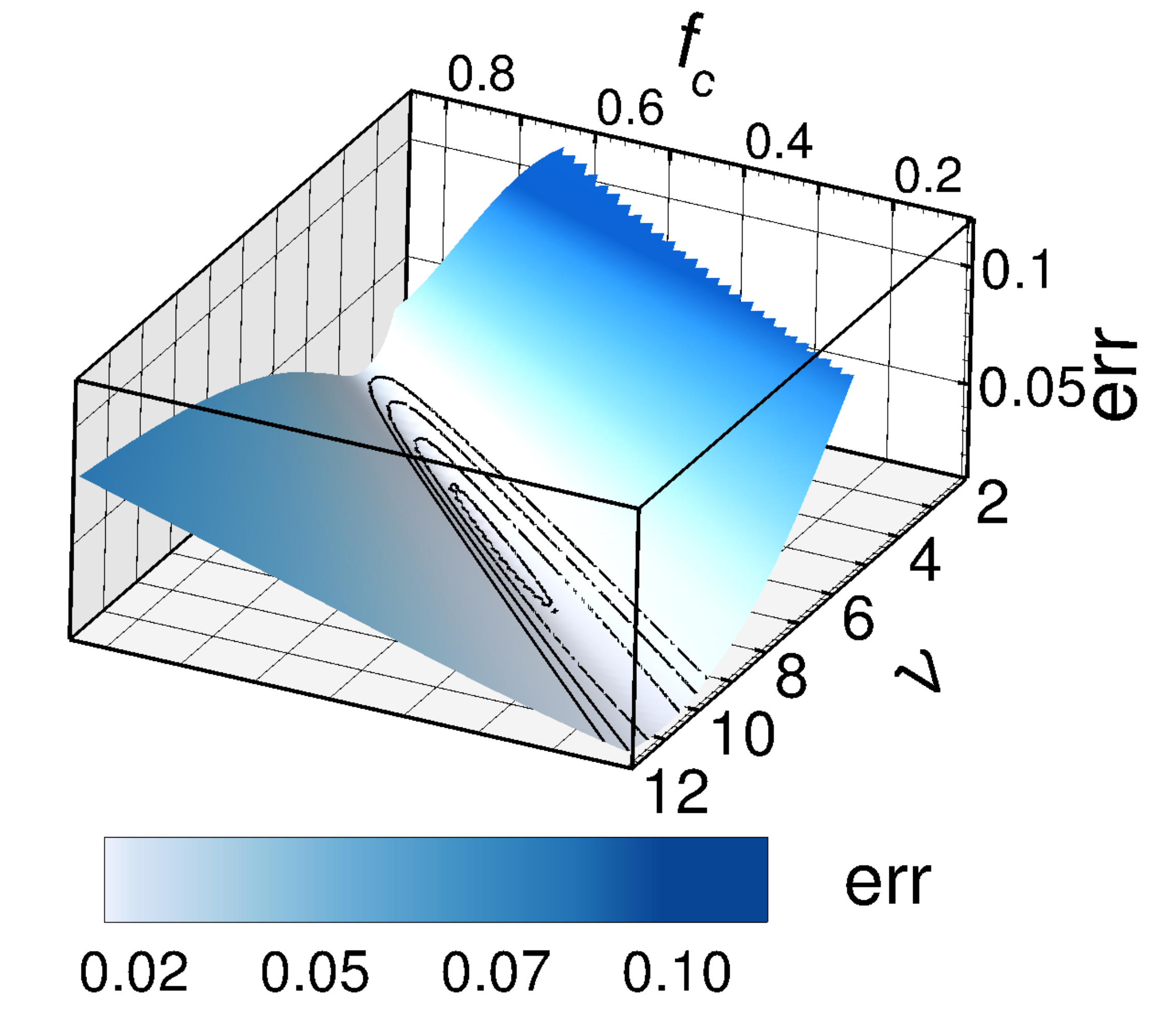}}\\
 \subfloat[]{\includegraphics[width=2.5in]{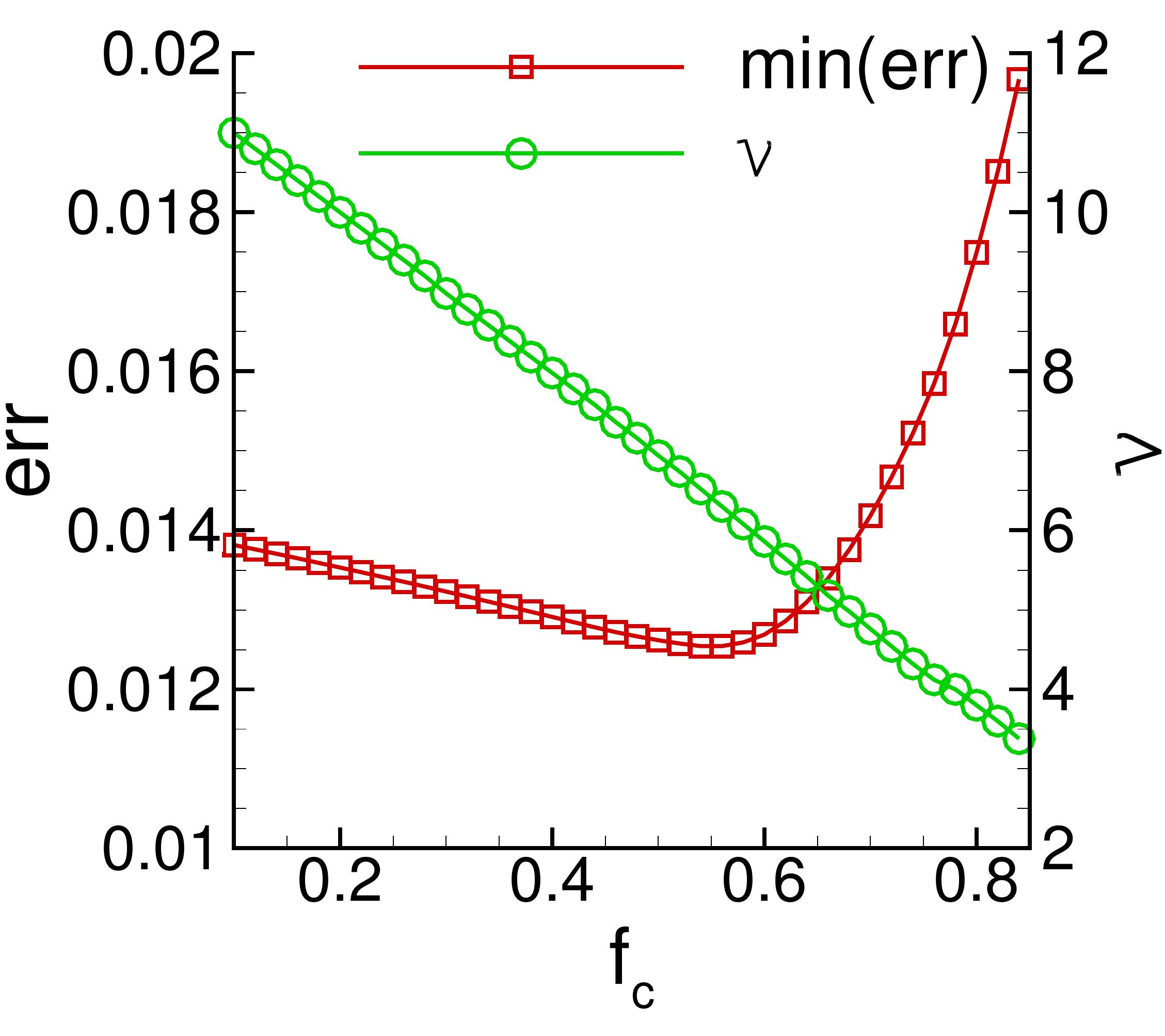}}
 \caption{Error plot for the $r_p=0.0,\mu=0.0$ system at $\rho=0.9$: (a) $err$ as a function of $f_c$ and $\nu$; the black lines are slices through different values of 
 $err$ and (b) minimum of $err$ (red line with squares) as a function of $f_c$; here for each $f_c$ value we choose $\nu$ that minimizes $err$; the used
values of $\nu$ are shown by the green line with circles. 
 Note different range compared to Fig.~\ref{err_plot_r02kt08}.}\label{err_plot_r00mu00}
\end{figure}

Next we proceed with discussing the scaling exponents for frictionless systems.   As an example, 
Fig.~\ref{err_plot_r00mu00} shows $err$ for the $r_p = 0.0,~\mu = 0.0$ system at $\rho = 0.9$ 
(Fig.~\ref{err_plot_r02kt08} shows the corresponding plots for the reference system).   Direct comparison
of these two figures shows the following: (i) the collapse appears to be much better for the considered
frictionless system (the minimum value of $err$ is smaller); (ii) the optimal values of $f_c,~\nu$ 
are significantly different: $\nu$ is much 
larger, and $f_c$ is much smaller for $r_p=0.0,~\mu=0.0$ system.   
We note that the minimum of $err$ curve in Fig.~\ref{err_plot_r00mu00}(b) is not as clearly 
defined as for the reference system, introducing some inaccuracy in the process of finding  optimal values of $f_c,~\nu$.  However, as it can be clearly seen in Fig.~\ref{err_plot_r00mu00}(b), 
this inaccuracy still limits $f_c$ to a very small value, $f_c < 0.6$, and
$\nu$ to a very large value, $\nu > 6$.     

Figure~\ref{err_plot_r00mu00} suggests some significant differences between $r_p = 0.0,~\mu = 0.0$ and
the reference system at $\rho = 0.9$.  
Table~\ref{fp_nu_frictionless} shows that the differences are present for other considered packing fractions
as well.    In particular, we always find large values of $\nu$ and very small values of $f_c$ for $r_p = 0.0,~\mu = 0.0$
system.  Small overall values of $err$ are a sign of a good quality of the collapse.  We note again that for the smallest $\rho$, the optimal
values of $f_c$ and $\nu$ are different from the rest, suggesting that scaling properties of force networks very
close to jamming transition may differ.   

One obvious question to ask is what causes a particularly large difference between $f_c$ and $f_p$ for $r_p = 0.0,~\mu = 0.0$ system (recall that typically  $f_p \in [1.05,1.25]$).   One possibility is that it may 
be caused by a finite system size.  Note that  the 
percolation threshold in a finite size system, $f_p$, is related to the one of the infinite size 
system, $f_p^\infty$, by~\cite{stauffer}
\begin{equation}
\vert f_p - f_p^{\infty} \vert \sim L^{-1/\nu}
\label{eq_Favg_vs_L}
\end{equation}
and $f_p^\infty < f_p$.   This relation suggests a strong influence of the system size on $f_p$ for large values of $\nu$,
such as those we are reporting in Table~\ref{fp_nu_frictionless} for $r_p = 0.0,~\mu = 0.0$ system.  Therefore, 
our conjecture is that the agreement of $f_c$ and $f_p$ could still be found if very large system size are considered.
In particular, if we assume that $f_c$ is close to $f_p^{\infty}$, the values of $f_c,~f_p$ would be for a very large system both in the range  $[0.5,0.7]$ for $r_p = 0.0,~\mu = 0.0$ system (see Table~\ref{fp_nu_frictionless}).

We close this section by pointing out that $\nu$ could in principle be computed using alternative approaches.   One avenue is to use 
Eq.~(\ref{eq_Favg_vs_L}); the value of $f_p$ is found as 
an average percolation threshold for each $L$ and plotted against the natural logarithm $\log (L^{-1/\nu})$; the slope of the linear fit 
should correspond to $-1/\nu$.  However, we find that the error of the linear fit is large, leading to the results that are less accurate
than the ones already obtained.  Alternatively, we could  estimate $\phi$ from the Fisher exponent $\tau$, see Eq.~(\ref{scaling_ns}), and 
the relation $\phi = (3-\tau)/(\tau-1)$. The results for $\phi$ obtained in this manner are again characterized by large error bars. 
We note that while both of the outlined approaches lead to the results that are inaccurate, they are still consistent with the ones found by scaling.

\subsection{Physical experiments: $\phi$ and the fractal dimension}\label{sec:exp}

In this section we report the results of physical experiments carried out with photoelastic particles, 
made from the PSM-1 sheets obtained from Vishay Precision Group; details about the material properties of 
these particles could be found in~\cite{clark_prl15}.   Figure~\ref{exp} shows the experimental setup
that consists of two plexiglass plates with a thin gap in between.  The size of the gap is slightly larger than the thickness of the particles.  
The domain is bounded by four walls, one of which is removable and  can slide in and out.   The experimental protocol consists of 
placing the particles in the gap, mixing them up, and than replacing the removable wall and gently applying desired pressure by a certain 
number ($1$ - $5$) of rubber bands.    The applied pressures lead only to modest inter-particle forces.    For each pressure, $5$ 
realizations are carried out.

The stress on the particles is visualized using cross-polarizers (see Fig.~\ref{exp}(b)). The photographs are processed using the Hough 
Transform~\cite{Duda} image 
processing technique to detect particles. From the brightness of the particles, the total stresses are computed via
$G^2$ method used extensively by Behringer's group (see, e.g.~\cite{clark_prl12}).  
The experiments are carried out using three particle sizes of diameters $0.58,~0.46$ and $0.41$ cm.  
We consider  a monodisperse system (with medium particles only) and two bidisperse ones that  
use large/medium and large/small particles.    For bidisperse systems, we always use equal area fraction
of particles of different sizes. Approximately $1,000$ particles are used in total.  

The obtained data are processed similarly to the ones resulting from the simulations, with the difference that here we focus on
the magnitude of the total stress on a particle, instead on contact forces, as in simulations.     Since only a single domain size is available, 
the domain is divided into $4,~8$ and $16$ smaller sub-domains of $1/4, 1/8$ and $1/16$ of the original  domain size. $D_f$ and $\phi$ 
are then computed using the box-counting method and by fitting the peaks of $S$-curves, respectively.
In what follows we focus on these two quantities only, since they could be obtained with a reasonable accuracy using available resources.   

\begin{figure}[t!]
 \subfloat[]{\includegraphics[width=2in]{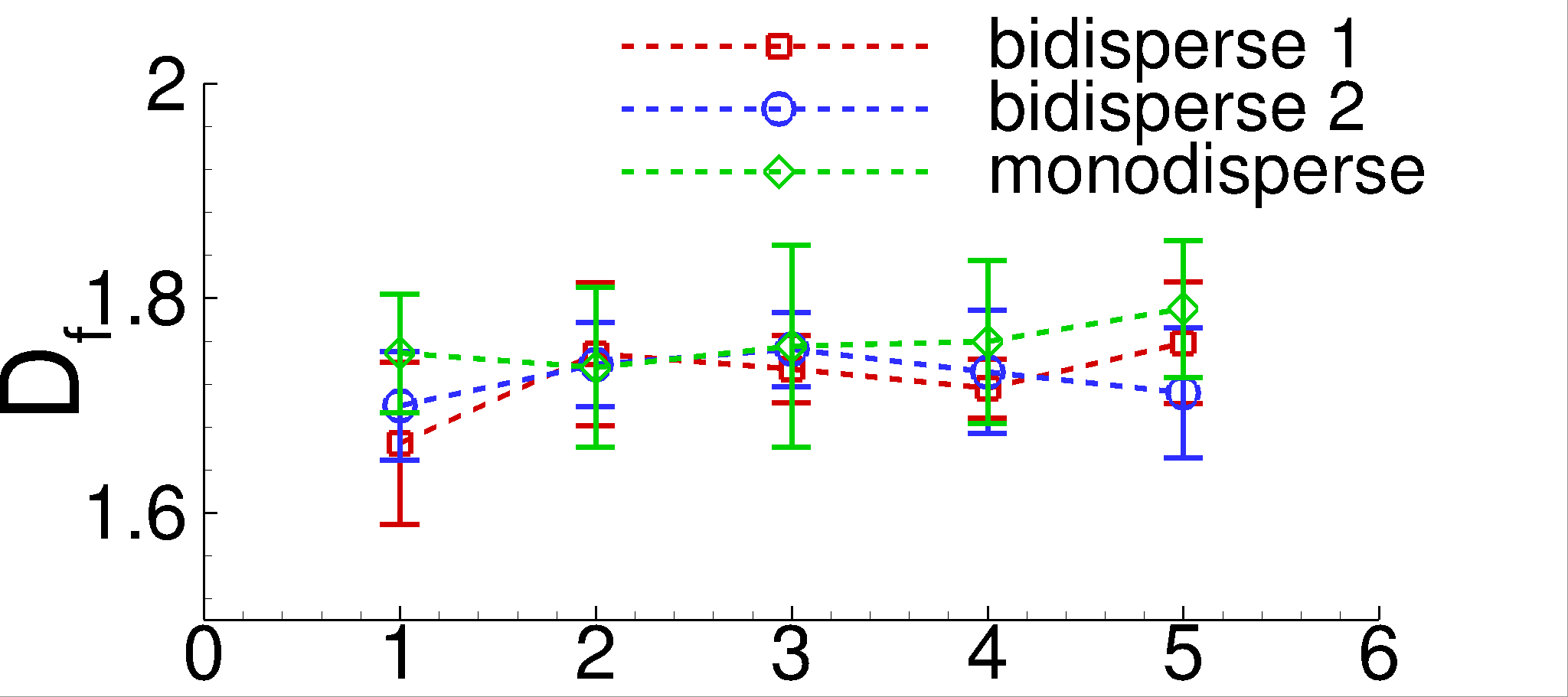}}\\
 \subfloat[]{\includegraphics[width=2in]{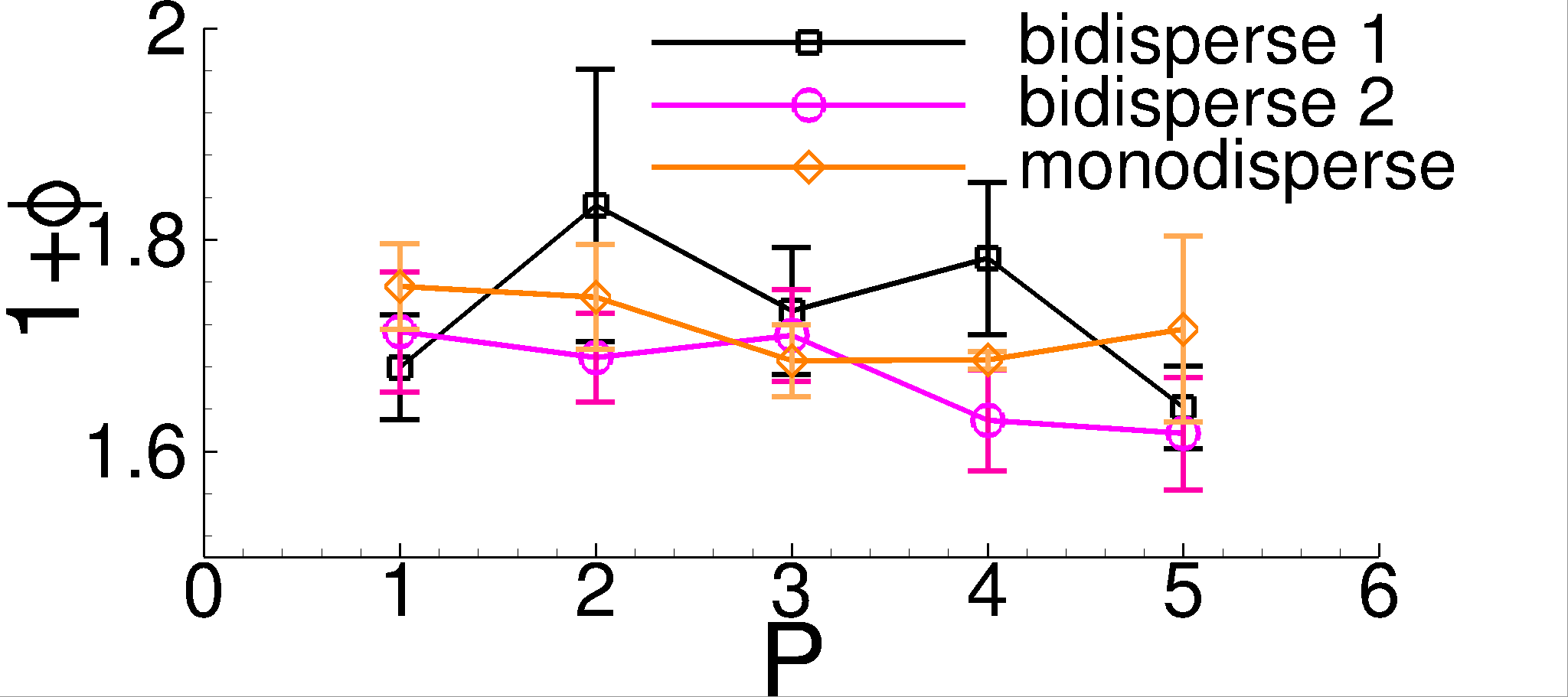}}
 \caption{(a) Fractal dimension, $D_f$, and (b) $1+\phi$ obtained from experiments carried out with 
 photoelastic particles, as a function of the applied pressure.   Bidisperse $1$ and $2$ refer to the
 systems of large/medium and large/small particles, respectively (see the text).  
}\label{fig_Df_fc}
\end{figure}

Figure~\ref{fig_Df_fc} shows $D_f$ and $1+\phi$ computed from the experimental data.   We note that while $D_f$ is consistent for the whole 
range of pressures applied and for all experimental setups, the value of $1+\phi$ has a larger variation, similarly as for the results obtained from the simulations.   
Since the number of realizations used for the experiments is much smaller, relatively large standard error is observed in the results. 
Still,  the experiments yield $D_f$ and $\phi$ that are consistent with the results obtained from the simulations 
carried out with frictional particles.    This is encouraging, in particular since the protocols in simulations and experiments differ 
(e.g., controlled pressure versus controlled packing fraction, additional friction between the particles and the substrate in experiments, 
that is not present in simulations); in addition, we have not attempted to precisely match the simulation parameters with material properties of 
the particles.   The consistency of the results therefore suggests that  they are independent of the protocol and of the material properties, at least
for the applied pressures considered.
We emphasize in particular that both simulation and experimental results lead to $D_f$ and $\phi$ that are 
significantly smaller than  the previously proposed value of $\phi \approx 0.9$~\cite{ostojic}.

\section{Conclusions}\label{sec:conclusions}

In this paper we focus on the scaling properties of force networks in compressed particulate systems in 
two spatial dimensions.    To complement
the results obtained by exploring scaling properties of the force networks, we also calculate the fractal 
dimension.   For disordered systems, we find that the scaling 
exponent, $\phi$, and the fractal dimension, $D_f$, are consistent over a range of considered packing fractions, $\rho$. 
The computed values are however significantly lower than the previously proposed ones, and in particular we find
that  $D_f \approx 1.8$.  This value is consistent with the one extracted from the physical 
experiments involving two dimensional systems of monodisperse and bidisperse frictional particles exposed 
to compression.   

Another significant finding is that the results for $\phi$ and $D_f$ are not consistent for the systems 
of monodispserse frictionless particles that partially crystallize under compression.   We obtain considerably different values of 
$D_f$ for these systems and present evidence that formation of partially ordered polycrystalline structure is 
the reason for this differences.   In addition, for these systems, we find that the values of the other scaling 
exponent, $\nu$, is very large while critical force threshold, $f_c$, is very small, when compared to the systems
that are not partially crystallized. Our results suggest that the particulate 
systems that develop ordered structures do not belong to the same (if any) universality class as the 
disordered ones.   

These results open new directions of research, including working towards understanding the conditions under which 
scaling properties of force networks are at least consistent if not universal.   Another question is how do 
our findings extend to three dimensional systems.   And finally, how the scaling properties of the force
networks relate to the macroscopic properties of the underlying physical systems.

{\it Acknowledgments}   The authors thank Robert Behringer and Mark Shattuck for the help with the experimental setup and
many useful discussions.   The experimental results were obtained and processed by a group of undergraduate students
(class of 2012) supervised by Te-Sheng Lin and the authors.   
This work was supported by the NSF Grants No. DMS-0835611 and DMS-1521717.

\bibliographystyle{apsrev} 

\begin{thebibliography}{33}
\expandafter\ifx\csname natexlab\endcsname\relax\def\natexlab#1{#1}\fi
\expandafter\ifx\csname bibnamefont\endcsname\relax
  \def\bibnamefont#1{#1}\fi
\expandafter\ifx\csname bibfnamefont\endcsname\relax
  \def\bibfnamefont#1{#1}\fi
\expandafter\ifx\csname citenamefont\endcsname\relax
  \def\citenamefont#1{#1}\fi
\expandafter\ifx\csname url\endcsname\relax
  \def\url#1{\texttt{#1}}\fi
\expandafter\ifx\csname urlprefix\endcsname\relax\def\urlprefix{URL }\fi
\providecommand{\bibinfo}[2]{#2}
\providecommand{\eprint}[2][]{\url{#2}}

\bibitem[{\citenamefont{Albert and Barab\'asi}(2002)}]{albert_barabasi_02}
\bibinfo{author}{\bibfnamefont{R.}~\bibnamefont{Albert}} \bibnamefont{and}
  \bibinfo{author}{\bibfnamefont{A.-L.} \bibnamefont{Barab\'asi}},
  \bibinfo{journal}{Rev. Mod. Phys.} \textbf{\bibinfo{volume}{74}},
  \bibinfo{pages}{47} (\bibinfo{year}{2002}).

\bibitem[{\citenamefont{Alexander}(1998)}]{alexander_physrep05}
\bibinfo{author}{\bibfnamefont{S.}~\bibnamefont{Alexander}},
  \bibinfo{journal}{Phys. Rep.} \textbf{\bibinfo{volume}{296}},
  \bibinfo{pages}{65} (\bibinfo{year}{1998}).

\bibitem[{\citenamefont{Mueth et~al.}(1998)\citenamefont{Mueth, Jaeger, and
  Nagel}}]{mueth98}
\bibinfo{author}{\bibfnamefont{D.~M.} \bibnamefont{Mueth}},
  \bibinfo{author}{\bibfnamefont{H.~M.} \bibnamefont{Jaeger}},
  \bibnamefont{and} \bibinfo{author}{\bibfnamefont{S.~R.} \bibnamefont{Nagel}},
  \bibinfo{journal}{Phys. Rev. E.} \textbf{\bibinfo{volume}{57}},
  \bibinfo{pages}{3164} (\bibinfo{year}{1998}).

\bibitem[{\citenamefont{Radjai et~al.}(1999)\citenamefont{Radjai, Moreau, and
  Roux}}]{radjai99}
\bibinfo{author}{\bibfnamefont{F.}~\bibnamefont{Radjai}},
  \bibinfo{author}{\bibfnamefont{J.~J.} \bibnamefont{Moreau}},
  \bibnamefont{and} \bibinfo{author}{\bibfnamefont{S.}~\bibnamefont{Roux}},
  \bibinfo{journal}{Chaos} \textbf{\bibinfo{volume}{9}}, \bibinfo{pages}{544}
  (\bibinfo{year}{1999}).

\bibitem[{\citenamefont{Radjai et~al.}(1998)\citenamefont{Radjai, Wolf, Jean,
  and Moreau}}]{radjai98b}
\bibinfo{author}{\bibfnamefont{F.}~\bibnamefont{Radjai}},
  \bibinfo{author}{\bibfnamefont{D.~E.} \bibnamefont{Wolf}},
  \bibinfo{author}{\bibfnamefont{M.}~\bibnamefont{Jean}}, \bibnamefont{and}
  \bibinfo{author}{\bibfnamefont{J.-J.} \bibnamefont{Moreau}},
  \bibinfo{journal}{Phys. Rev. Lett.} \textbf{\bibinfo{volume}{80}},
  \bibinfo{pages}{61} (\bibinfo{year}{1998}).

\bibitem[{\citenamefont{Majmudar and Behringer}(2005)}]{majmudar05a}
\bibinfo{author}{\bibfnamefont{T.~S.} \bibnamefont{Majmudar}} \bibnamefont{and}
  \bibinfo{author}{\bibfnamefont{R.~P.} \bibnamefont{Behringer}},
  \bibinfo{journal}{Nature} \textbf{\bibinfo{volume}{435}},
  \bibinfo{pages}{1079} (\bibinfo{year}{2005}).

\bibitem[{\citenamefont{Silbert}(2006)}]{silbert_pre06}
\bibinfo{author}{\bibfnamefont{L.~E.} \bibnamefont{Silbert}},
  \bibinfo{journal}{Phys. Rev. E} \textbf{\bibinfo{volume}{74}},
  \bibinfo{pages}{051303} (\bibinfo{year}{2006}).

\bibitem[{\citenamefont{Wambaugh}(2010)}]{Wambaugh_physD_10}
\bibinfo{author}{\bibfnamefont{J.}~\bibnamefont{Wambaugh}},
  \bibinfo{journal}{Phys. D} \textbf{\bibinfo{volume}{239}},
  \bibinfo{pages}{1818} (\bibinfo{year}{2010}).

\bibitem[{\citenamefont{Peters et~al.}(2005)\citenamefont{Peters, Muthuswamy,
  Wibowo, and Tordesillas}}]{peters05}
\bibinfo{author}{\bibfnamefont{J.}~\bibnamefont{Peters}},
  \bibinfo{author}{\bibfnamefont{M.}~\bibnamefont{Muthuswamy}},
  \bibinfo{author}{\bibfnamefont{J.}~\bibnamefont{Wibowo}}, \bibnamefont{and}
  \bibinfo{author}{\bibfnamefont{A.}~\bibnamefont{Tordesillas}},
  \bibinfo{journal}{Phys. Rev. E} \textbf{\bibinfo{volume}{72}},
  \bibinfo{pages}{041307} (\bibinfo{year}{2005}).

\bibitem[{\citenamefont{Tordesillas et~al.}(2010)\citenamefont{Tordesillas,
  Walker, and Lin}}]{tordesillas_pre10}
\bibinfo{author}{\bibfnamefont{A.}~\bibnamefont{Tordesillas}},
  \bibinfo{author}{\bibfnamefont{D.~M.} \bibnamefont{Walker}},
  \bibnamefont{and} \bibinfo{author}{\bibfnamefont{Q.}~\bibnamefont{Lin}},
  \bibinfo{journal}{Phys. Rev. E} \textbf{\bibinfo{volume}{81}},
  \bibinfo{pages}{011302} (\bibinfo{year}{2010}).

\bibitem[{\citenamefont{Tordesillas et~al.}(2012)\citenamefont{Tordesillas,
  Walker, Froyland, Zhang, and Behringer}}]{tordesillas_bob_pre12}
\bibinfo{author}{\bibfnamefont{A.}~\bibnamefont{Tordesillas}},
  \bibinfo{author}{\bibfnamefont{D.~M.} \bibnamefont{Walker}},
  \bibinfo{author}{\bibfnamefont{G.}~\bibnamefont{Froyland}},
  \bibinfo{author}{\bibfnamefont{J.}~\bibnamefont{Zhang}}, \bibnamefont{and}
  \bibinfo{author}{\bibfnamefont{R.}~\bibnamefont{Behringer}},
  \bibinfo{journal}{Phys. Rev. E} \textbf{\bibinfo{volume}{86}},
  \bibinfo{pages}{011306} (\bibinfo{year}{2012}).

\bibitem[{\citenamefont{Bassett et~al.}(2012)\citenamefont{Bassett, Owens,
  Daniels, and Porter}}]{daniels_pre12}
\bibinfo{author}{\bibfnamefont{D.~S.} \bibnamefont{Bassett}},
  \bibinfo{author}{\bibfnamefont{E.~T.} \bibnamefont{Owens}},
  \bibinfo{author}{\bibfnamefont{K.~E.} \bibnamefont{Daniels}},
  \bibnamefont{and} \bibinfo{author}{\bibfnamefont{M.~A.}
  \bibnamefont{Porter}}, \bibinfo{journal}{Phys. Rev. E}
  \textbf{\bibinfo{volume}{86}}, \bibinfo{pages}{041306}
  (\bibinfo{year}{2012}).

\bibitem[{\citenamefont{Herrera et~al.}(2011)\citenamefont{Herrera, McCarthy,
  Sotterbeck, Cephas, Losert, and Girvan}}]{herrera_pre11}
\bibinfo{author}{\bibfnamefont{M.}~\bibnamefont{Herrera}},
  \bibinfo{author}{\bibfnamefont{S.}~\bibnamefont{McCarthy}},
  \bibinfo{author}{\bibfnamefont{S.}~\bibnamefont{Sotterbeck}},
  \bibinfo{author}{\bibfnamefont{E.}~\bibnamefont{Cephas}},
  \bibinfo{author}{\bibfnamefont{W.}~\bibnamefont{Losert}}, \bibnamefont{and}
  \bibinfo{author}{\bibfnamefont{M.}~\bibnamefont{Girvan}},
  \bibinfo{journal}{Phys. Rev. E} \textbf{\bibinfo{volume}{83}},
  \bibinfo{pages}{061303} (\bibinfo{year}{2011}).

\bibitem[{\citenamefont{Walker and Tordesillas}(2012)}]{walker_pre12}
\bibinfo{author}{\bibfnamefont{D.}~\bibnamefont{Walker}} \bibnamefont{and}
  \bibinfo{author}{\bibfnamefont{A.}~\bibnamefont{Tordesillas}},
  \bibinfo{journal}{Phys. Rev. E} \textbf{\bibinfo{volume}{85}},
  \bibinfo{pages}{011304} (\bibinfo{year}{2012}).

\bibitem[{\citenamefont{Ar\'evalo et~al.}(2010)\citenamefont{Ar\'evalo,
  Zuriguel, and Maza}}]{arevalo_pre10}
\bibinfo{author}{\bibfnamefont{R.}~\bibnamefont{Ar\'evalo}},
  \bibinfo{author}{\bibfnamefont{I.}~\bibnamefont{Zuriguel}}, \bibnamefont{and}
  \bibinfo{author}{\bibfnamefont{D.}~\bibnamefont{Maza}},
  \bibinfo{journal}{Phys. Rev. E} \textbf{\bibinfo{volume}{81}},
  \bibinfo{pages}{041302} (\bibinfo{year}{2010}).

\bibitem[{\citenamefont{Ar\'evalo et~al.}(2013)\citenamefont{Ar\'evalo,
  Pugnaloni, Zuriguel, and Maza}}]{arevalo_pre13}
\bibinfo{author}{\bibfnamefont{R.}~\bibnamefont{Ar\'evalo}},
  \bibinfo{author}{\bibfnamefont{L.~A.} \bibnamefont{Pugnaloni}},
  \bibinfo{author}{\bibfnamefont{I.}~\bibnamefont{Zuriguel}}, \bibnamefont{and}
  \bibinfo{author}{\bibfnamefont{D.}~\bibnamefont{Maza}},
  \bibinfo{journal}{Phys. Rev. E} \textbf{\bibinfo{volume}{87}},
  \bibinfo{pages}{022203} (\bibinfo{year}{2013}).

\bibitem[{\citenamefont{Kondic et~al.}(2012)\citenamefont{Kondic, Goullet,
  O'Hern, Kramar, Mischaikow, and Behringer}}]{epl12}
\bibinfo{author}{\bibfnamefont{L.}~\bibnamefont{Kondic}},
  \bibinfo{author}{\bibfnamefont{A.}~\bibnamefont{Goullet}},
  \bibinfo{author}{\bibfnamefont{C.}~\bibnamefont{O'Hern}},
  \bibinfo{author}{\bibfnamefont{M.}~\bibnamefont{Kramar}},
  \bibinfo{author}{\bibfnamefont{K.}~\bibnamefont{Mischaikow}},
  \bibnamefont{and}
  \bibinfo{author}{\bibfnamefont{R.}~\bibnamefont{Behringer}},
  \bibinfo{journal}{Europhys. Lett.} \textbf{\bibinfo{volume}{97}},
  \bibinfo{pages}{54001} (\bibinfo{year}{2012}).

\bibitem[{\citenamefont{Kramar et~al.}(2013)\citenamefont{Kramar, Goullet,
  Kondic, and Mischaikow}}]{pre13}
\bibinfo{author}{\bibfnamefont{M.}~\bibnamefont{Kramar}},
  \bibinfo{author}{\bibfnamefont{A.}~\bibnamefont{Goullet}},
  \bibinfo{author}{\bibfnamefont{L.}~\bibnamefont{Kondic}}, \bibnamefont{and}
  \bibinfo{author}{\bibfnamefont{K.}~\bibnamefont{Mischaikow}},
  \bibinfo{journal}{Phys. Rev. E} \textbf{\bibinfo{volume}{87}},
  \bibinfo{pages}{042207} (\bibinfo{year}{2013}).

\bibitem[{\citenamefont{Kramar et~al.}(2014)\citenamefont{Kramar, Goullet,
  Kondic, and Mischaikow}}]{pre14}
\bibinfo{author}{\bibfnamefont{M.}~\bibnamefont{Kramar}},
  \bibinfo{author}{\bibfnamefont{A.}~\bibnamefont{Goullet}},
  \bibinfo{author}{\bibfnamefont{L.}~\bibnamefont{Kondic}}, \bibnamefont{and}
  \bibinfo{author}{\bibfnamefont{K.}~\bibnamefont{Mischaikow}},
  \bibinfo{journal}{Phys. Rev. E} \textbf{\bibinfo{volume}{90}},
  \bibinfo{pages}{052203} (\bibinfo{year}{2014}).

\bibitem[{\citenamefont{Ostojic et~al.}(2006)\citenamefont{Ostojic, Somfai, and
  Nienhuis}}]{ostojic}
\bibinfo{author}{\bibfnamefont{S.}~\bibnamefont{Ostojic}},
  \bibinfo{author}{\bibfnamefont{E.}~\bibnamefont{Somfai}}, \bibnamefont{and}
  \bibinfo{author}{\bibfnamefont{B.}~\bibnamefont{Nienhuis}},
  \bibinfo{journal}{Nature} \textbf{\bibinfo{volume}{439}},
  \bibinfo{pages}{828} (\bibinfo{year}{2006}).

\bibitem[{\citenamefont{Ostojic et~al.}(2007)\citenamefont{Ostojic, Vlugt, and
  Nienhuis}}]{ostojic_pre07}
\bibinfo{author}{\bibfnamefont{S.}~\bibnamefont{Ostojic}},
  \bibinfo{author}{\bibfnamefont{T.}~\bibnamefont{Vlugt}}, \bibnamefont{and}
  \bibinfo{author}{\bibfnamefont{B.}~\bibnamefont{Nienhuis}},
  \bibinfo{journal}{Phys. Rev. E} \textbf{\bibinfo{volume}{75}}
  (\bibinfo{year}{2007}).

\bibitem[{\citenamefont{Pastor-Satorras and Miguel}(2012)}]{Pastor-Satorras_12}
\bibinfo{author}{\bibfnamefont{R.}~\bibnamefont{Pastor-Satorras}}
  \bibnamefont{and} \bibinfo{author}{\bibfnamefont{M.-C.}
  \bibnamefont{Miguel}}, \bibinfo{journal}{J. of Stat. Mech.}
  \textbf{\bibinfo{volume}{2012}} (\bibinfo{year}{2012}).

\bibitem[{\citenamefont{Stauffer and Aharony}(1994)}]{stauffer}
\bibinfo{author}{\bibfnamefont{D.}~\bibnamefont{Stauffer}} \bibnamefont{and}
  \bibinfo{author}{\bibfnamefont{A.}~\bibnamefont{Aharony}},
  \emph{\bibinfo{title}{Introduction to Percolation Theory}}
  (\bibinfo{publisher}{CRC Press}, \bibinfo{year}{1994}).

\bibitem[{\citenamefont{Kovalcinova et~al.}(2015)\citenamefont{Kovalcinova,
  Goullet, and Kondic}}]{kovalcinova_15}
\bibinfo{author}{\bibfnamefont{L.}~\bibnamefont{Kovalcinova}},
  \bibinfo{author}{\bibfnamefont{A.}~\bibnamefont{Goullet}}, \bibnamefont{and}
  \bibinfo{author}{\bibfnamefont{L.}~\bibnamefont{Kondic}},
  \bibinfo{journal}{Phys. Rev. E} \textbf{\bibinfo{volume}{92}},
  \bibinfo{pages}{032204} (\bibinfo{year}{2015}).

\bibitem[{\citenamefont{Geng et~al.}(2003)\citenamefont{Geng, Behringer,
  Reydellet, and Cl\'{e}ment}}]{geng_physicad03}
\bibinfo{author}{\bibfnamefont{J.}~\bibnamefont{Geng}},
  \bibinfo{author}{\bibfnamefont{R.~P.} \bibnamefont{Behringer}},
  \bibinfo{author}{\bibfnamefont{G.}~\bibnamefont{Reydellet}},
  \bibnamefont{and}
  \bibinfo{author}{\bibfnamefont{E.}~\bibnamefont{Cl\'{e}ment}},
  \bibinfo{journal}{Phys. D} \textbf{\bibinfo{volume}{182}},
  \bibinfo{pages}{274} (\bibinfo{year}{2003}).

\bibitem[{\citenamefont{Kondic}(1999)}]{kondic_99}
\bibinfo{author}{\bibfnamefont{L.}~\bibnamefont{Kondic}},
  \bibinfo{journal}{Phys. Rev. E} \textbf{\bibinfo{volume}{60}},
  \bibinfo{pages}{751} (\bibinfo{year}{1999}).

\bibitem[{\citenamefont{Cundall and Strack}(1979)}]{cundall79}
\bibinfo{author}{\bibfnamefont{P.~A.} \bibnamefont{Cundall}} \bibnamefont{and}
  \bibinfo{author}{\bibfnamefont{O.~D.~L.} \bibnamefont{Strack}},
  \bibinfo{journal}{G\'eotechnique} \textbf{\bibinfo{volume}{29}},
  \bibinfo{pages}{47} (\bibinfo{year}{1979}).

\bibitem[{\citenamefont{Brendel and Dippel}(1998)}]{brendel98}
\bibinfo{author}{\bibfnamefont{L.}~\bibnamefont{Brendel}} \bibnamefont{and}
  \bibinfo{author}{\bibfnamefont{S.}~\bibnamefont{Dippel}}, in
  \emph{\bibinfo{booktitle}{Physics of Dry Granular Media}}, edited by
  \bibinfo{editor}{\bibfnamefont{H.~J.} \bibnamefont{Herrmann}},
  \bibinfo{editor}{\bibfnamefont{J.-P.} \bibnamefont{Hovi}}, \bibnamefont{and}
  \bibinfo{editor}{\bibfnamefont{S.}~\bibnamefont{Luding}}
  (\bibinfo{publisher}{Kluwer Academic Publishers},
  \bibinfo{address}{Dordrecht}, \bibinfo{year}{1998}), p. \bibinfo{pages}{313}.

\bibitem[{\citenamefont{Goldenberg and Goldhirsch}(2005)}]{goldhirsch_nature05}
\bibinfo{author}{\bibfnamefont{C.}~\bibnamefont{Goldenberg}} \bibnamefont{and}
  \bibinfo{author}{\bibfnamefont{I.}~\bibnamefont{Goldhirsch}},
  \bibinfo{journal}{Nature} \textbf{\bibinfo{volume}{435}},
  \bibinfo{pages}{188} (\bibinfo{year}{2005}).

\bibitem[{\citenamefont{Truskett et~al.}(1998)\citenamefont{Truskett, Torquato,
  Sastry, Debenedetti, and Stillinger}}]{torquato_98}
\bibinfo{author}{\bibfnamefont{T.~M.} \bibnamefont{Truskett}},
  \bibinfo{author}{\bibfnamefont{S.}~\bibnamefont{Torquato}},
  \bibinfo{author}{\bibfnamefont{S.}~\bibnamefont{Sastry}},
  \bibinfo{author}{\bibfnamefont{P.~G.} \bibnamefont{Debenedetti}},
  \bibnamefont{and} \bibinfo{author}{\bibfnamefont{F.~H.}
  \bibnamefont{Stillinger}}, \bibinfo{journal}{Phys. Rev. E}
  \textbf{\bibinfo{volume}{58}}, \bibinfo{pages}{3083} (\bibinfo{year}{1998}).

\bibitem[{\citenamefont{Clark et~al.}(2015)\citenamefont{Clark, Petersen,
  Kondic, and Behringer}}]{clark_prl15}
\bibinfo{author}{\bibfnamefont{A.~H.} \bibnamefont{Clark}},
  \bibinfo{author}{\bibfnamefont{A.~J.} \bibnamefont{Petersen}},
  \bibinfo{author}{\bibfnamefont{L.}~\bibnamefont{Kondic}}, \bibnamefont{and}
  \bibinfo{author}{\bibfnamefont{R.~P.} \bibnamefont{Behringer}},
  \bibinfo{journal}{Phys. Rev. Lett.} \textbf{\bibinfo{volume}{114}},
  \bibinfo{pages}{144502} (\bibinfo{year}{2015}).

\bibitem[{\citenamefont{Duda and Hart}(1972)}]{Duda}
\bibinfo{author}{\bibfnamefont{R.~O.} \bibnamefont{Duda}} \bibnamefont{and}
  \bibinfo{author}{\bibfnamefont{P.~E.} \bibnamefont{Hart}},
  \bibinfo{journal}{Commun. ACM} \textbf{\bibinfo{volume}{15}},
  \bibinfo{pages}{11} (\bibinfo{year}{1972}).

\bibitem[{\citenamefont{Clark et~al.}(2012)\citenamefont{Clark, Kondic, and
  Behringer}}]{clark_prl12}
\bibinfo{author}{\bibfnamefont{A.~H.} \bibnamefont{Clark}},
  \bibinfo{author}{\bibfnamefont{L.}~\bibnamefont{Kondic}}, \bibnamefont{and}
  \bibinfo{author}{\bibfnamefont{R.~P.} \bibnamefont{Behringer}},
  \bibinfo{journal}{Phys. Rev. Lett.} \textbf{\bibinfo{volume}{109}},
  \bibinfo{pages}{238302} (\bibinfo{year}{2012}).

\end{thebibliography}

\end{document}